\documentclass[aps,pre,floatfix,twocolumn,showpacs,nofootinbib]{revtex4}
\usepackage{graphicx} \usepackage{amsmath} \usepackage{amssymb} 
\usepackage[sort&compress]{natbib}

\newcommand{\BEA}{\begin{eqnarray}}
\newcommand{\EEA}{\end{eqnarray}}
\newcommand{\bq}{\begin{eqnarray}}
\newcommand{\be}{\begin{equation}}
\newcommand{\eq}{\end{eqnarray}}
\newcommand{\ee}{\end{equation}}

\newcommand{\ssum}{{\sum} }
\newcommand{\pprod}{{\prod} }

\renewcommand{\d}{{\rm d}}
\newcommand{\x}{ {\bf x}}
\newcommand{\y}{ {\bf y}}
\newcommand{\pr}{ \prime}
\renewcommand{\i}{\imath}
\newcommand{\e}{\mathfrak{e}\,}
\newcommand{\m}{\mathfrak{m}\,}
\newcommand{\CZ}{{\cal Z}}
\newcommand{\CP}{{\cal P}}
\newcommand{\ep}{{\epsilon}}
\newcommand{\pd}{\partial}
\newcommand{\G}{G}
\newcommand{\nn}{\nonumber}
\newcommand{\comment}[1]{}
\def\dbarrm {{\mathchar'26\mkern-11mu{\rm d}}}                        %
\begin{document} 

\draft
\title{Thermodynamic efficiency of information and heat flow}

\author{ Armen E. Allahverdyan$^1$, Dominik Janzing$^2$, Guenter Mahler$^3$}
\affiliation{$^1$Yerevan Physics Institute,
Alikhanian Brothers Street 2, Yerevan 375036, Armenia,\\
$^2$ MPI for Biological Cybernetics, Spemannstrasse 38
72076 Tuebingen, Germany,
$^3$Institute of Theoretical Physics I, University of Stuttgart, 
Pfaffenwaldring 57, 70550 Stuttgart, Germany}

\begin{abstract} A basic task of information processing is information
transfer (flow). Here we study a pair of Brownian particles each coupled
to a thermal bath at temperature $T_1$ and $T_2$, respectively.  The
information flow in such a system is defined via the time-shifted mutual information.
The information
flow nullifies at equilibrium, and its efficiency is defined as the ratio of
flow over the total entropy production in the system. For a
stationary state the information flows from higher to lower
temperatures, and its the efficiency is bound from above by $\frac{{\rm
max}[T_1,T_2]}{|T_1-T_2|}$. This upper bound is imposed by the second
law and it quantifies the thermodynamic cost for information flow in the
present class of systems.  It can be reached in the adiabatic situation,
where the particles have widely different characteristic times.  The
efficiency of heat flow|defined as the heat flow over the total amount
of dissipated heat|is limited from above by the same factor. There is a
complementarity between heat- and information-flow: the setup which is
most efficient for the former is the least efficient for the latter and
{\it vice versa}. The above bound for the efficiency can be
[transiently] overcome in certain non-stationary situations, but the
efficiency is still limited from above. We study yet another measure of
information-processing [transfer entropy] proposed in literature. Though
this measure does not require any thermodynamic cost, the information
flow and transfer entropy are shown to be intimately related for
stationary states. 

\end{abstract}


\comment{89.70.-a Information and communication theory 
89.70.Cf Entropy and other measures of information 
89.75.-k Complex systems 
02.50.-r Probability theory, stochastic processes, and statistics
05.10.Gg Stochastic analysis methods (Fokker-Planck, Langevin, etc.
}
\maketitle

\section{Introduction}

Relations between statistical thermodynamics and information science
have long since been recognized; they have been the source of mutual
fertilization, but occasionally also of confusion. Pertinent examples
are the nature of the maximum entropy principle \cite{max_entropy} and the analysis of
the Maxwell demon concept \cite{maxwell}. Then there is the huge field of
infomation processing devices. Their continuing miniaturization \cite{venturi} is
approaching already the nanometer scale, which makes it obligatory to
study the information carriers as physical entities subject to the laws
of statistical physics. Note that information processing constitutes a certain
class of tasks (functionality) to be implemented on a physical system.
Basic tasks of such kind are, {\it inter alia}, information erasure and
information transfer. 

The thermodynamic cost of information erasure received much attention
\cite{landauer_principle,shizume,Dominik,plenio,fahn,armen,ishioka,ueda_sagawa};
see \cite{plenio} for a review. The information erasure is governed by
the Landauer principle \cite{landauer_principle,shizume,Dominik,plenio},
which |together with its limitations
\cite{fahn,armen,ishioka,ueda_sagawa}|has been investigated from
different perspectives. 

The task of information transfer is not uniquely defined, but has to be specified before one can start any
detailed physical investigation (see below). Its thermodynamical cost in
the presence of a thermal bath has been studied in
\cite{landauer_communication,toffoli,benioff,bennett,bennett_brownian,porod,parker_walker,levitin}.
There is an essential difference between the problem of thermodynamic
costs during information erasure versus those during information
transfer. In the former case the information carrier has to be an open,
or even dissipative system, while for the latter case the external bath
frequently plays a role of a hindrance
\cite{bennett_brownian,porod,levitin}. For a finite, conservative system
the problem of thermodynamic costs is not well-defined, since for such
systems the proper measures of irreversibility and dissipation are
lacking; see, however, \cite{lindblad} in this context.  Thus, the
thermodynamic cost for information transfer is to be studied in specific
macroscopic settings.

The current status of the problem is controversial: in the literature
there are statements claiming both the existence \cite{porod,parker_walker,levitin}
and the absence
\cite{landauer_communication,toffoli,benioff,bennett,bennett_brownian}
of inevitable thermodynamic costs during information transfer.  The
arguments against the fundamental bounds on thermodynamic costs in
information transfer
\cite{landauer_communication,toffoli,benioff,bennett,bennett_brownian}
rely in essence on the known statistical physics fact that the entropy
generated during a process can be nullified by nullifying the rate of
this process \cite{ll}.  However, this and related arguments leave
unanswered the question on whether there are thermodynamic costs in a
more realistic case of finite-rate information transfer. In practice,
the information transfer should normally proceed at a finite rate. 

The presently known arguments in favour of the existence of
thermodynamic cost during information transfer are either heuristic
\cite{porod,parker_walker}, or concentrate on those aspects of computation and
communication, which require some energy for carrying out the task, but
this energy is not necessarily dissipated; see
\cite{levitin,bremermann,gea,gea_kish}
\footnote{The authors of \cite{gea_kish} point out that
although in principle the energy required for carrying out these tasks
need not be dissipated, in practice it is normally dissipated, at least
partially.}.
The present work aims at clarifying how processes of information and
heat flow relate to dissipative charateristics such as entropy
production or heat dissipation. In particular, we clarify which taks 
of information processing do (not) reqiure thermodynamic cost.											

We shall work with the|supposedly|simplest set-up that allows to study
the above problem: two (classical) Brownian particles each interacting
with a thermal bath. The model combines two basic ingredients needed for
studying the information transfer: randomness|which is necessary for the
very notion of information|and directedness, i.e., the possibility of
inducing a current of physical quantities via externally imposed
gradients of temperature and/or various potentials. 

For the present bi-partite problem, whose dynamics is formulated as a
Markov process for random coordinates $X_1(t)$ and $X_2(t)$ of Brownian
particle, the mutual information is given by the known Shannon
expression $I[X_1(t):X_2(t)]$ \cite{ash,strat_IT,cover} 
\footnote{We stress that the employed notion of information concerns its syntactic aspects and
does not concern its semantic [meaning] and pragmatic [purpose] aspects. Indeed, the problems we intend to 
study|physics of information carriers, thermodynamic cost of information flow, {\it etc}|refer
primarily to syntactic aspects. }. This is an
ensemble property, which is naturally symmetric
$I[X_1(t):X_2(t)]=I[X_2(t):X_1(t)]$, and has the same status as other
macroscopic observables in statistical physics, e.g., the average
energy. The information flow $\i_{2\to 1}$ is defined via
the time-shifted mutual
information via $\i_{2\to 1} =\pd_\tau I[X_1(t+\tau):X_2(t)]|_{\tau\to
+0}$. The full rate $\frac{\d}{\d t}I[X_1(t):X_2(t)]$ of mutual
information is now separated into the information that has flown from
the first to the second particle $\i_{1\to 2}$ and {\it vice versa}:
$\frac{\d}{\d t}I[X_1(t):X_2(t)]=\i_{1\to 2}+\i_{2\to 1}$.
We discuss this definition in section \ref{i_flow}, explain its
relation to prediction and clarify the meaning of a negative information
flow. The usage of the time-shifted mutual information for quantifying
the information flow was advocated in \cite{vastano,neuro_vastano,ho_vastano,nichols_vastano}. 

The information flow is an asymmetric quantity, $\i_{2\to
1}\not=\i_{1\to 2}$, and it allows to distinguish between the source of
information versus its recipient (such a distinction cannot be done via
the mutual information, which is a symmetric quantity). Another
important feature of information flow is that|for the considered
bi-partite Markovian system|it nullifies in equilibrium: $\i_{2\to
1}=\i_{1\to 2}=0$, because the information flow appears to be related to
the entropy flow; see section \ref{i_flow_thermo}. 

Once we have shown that the information flow is absent in equilibrium,
we concentrate on the case when this flow is induced by a temperature
gradient. Despite some formal similarities, a temperature gradient and a
potential gradient|two main sources of non-equiilbrium|are of different
physical origin and enjoy different features; see \cite{michel} for a
recent discussion. Hence we expect that the specific features of the
information flow will differ depending on the type of non-equilibrium
situation. In this context, the temperature-gradient situation is
perhaps the first one to study, since the function of the information
transfer comes close to the function of a thermodynamic machine.  This
far reaching analogy reflects itself in the features of the efficiency
of information transfer, which is defined as the useful product, i.e.,
the information flow, divided over the total waste, as quantified by the
entropy production in the overall system. In the stationary
two-temperature state, where both the information flow and entropy
production are time-independent, the efficiency of information flow is
limited from above by $\frac{{\rm max}[T_1,T_2]}{|T_1-T_2|}$. This
expression depends only on temperatures and does not depend on various
details of the system (such as damping constants, inter-particle
potentials, {\it etc}). The existence of such an upper bound implies a
definite thermodynamic cost for information flow. Interestingly, the
upper bound for the efficiency can be reached in the adiabatic
situation, when the source of information is much slower than its
recipient. This fact to some extent resembles the reachability of the
Carnot bound for thermodynamic efficiency of heat-engines and
refrigerators \cite{ll}. The upper bound $\frac{{\rm
max}[T_1,T_2]}{|T_1-T_2|}$ for the efficiency of information flow can be
well surpassed in certain non-stationary, transient situations. Even
then the efficiency of information flow is limited from above via the
physical parameters of the system. 

We shall argue that there is a clear parallel in the definition of
information flow and heat flow. This is additionally underlined by the
fact that in the stationary state both heat and information flow from
higher temperatures to lower ones. Moreover, the efficiency of heat
flow|defined accordingly as the ration of the heat flow and heat
dissipation rate|appears to be limited from above by the same factor
$\frac{{\rm max}[T_1,T_2]}{|T_1-T_2|}$. There is however an important
complementarity here: the upper bound for the efficiency of heat flow is
is reached exactly for that setup which is the worst one for the
efficiency of information flow; see section \ref{complo}. 

The information flow $\i_{2\to 1}$ characterizes the predictability of
(future of) $1$ (first particle) from the viewpoint of $2$ (second
particle). Another aspect of information processing in stochastic
systems concerns predicting by $1$ its own future, and the help provided
by $2$ in accomplishing this task. This is essentially the notion of
Granger-causality first proposed in econometrics for quantifying causal
relation between coupled stochastic processes
\cite{granger_general_definition,granger_causality}, and formalized in
information-theoretic terms via the concept of transfer entropy
\cite{marko,schreiber}. It appears that this type of information
processing does not require any thermodynamic cost, i.e., the transfer
entropy|in contrast to information flow|does not necessarily nullify at
equilibrium.  Despite of this, there are interesting relations between
the transfer entropy and information flow, which are partially uncovered
in section \ref{i_flow_t_entropy}. 

The paper is organized as follows. Section \ref{basic} reminds
the definition of entropy and mutual information; this reminder is
continued in Appendix \ref{info-meaning}.  Section \ref{model} defines
the class of models to be studied. In section \ref{i_flow} we discuss in
detail the information-theoretic definition of information flow. This
discussion is continued in section \ref{i_flow_thermo} with 
relating the information flow to the entropy flow. The same section recalls
the concepts of entropy production and heat dissipation. The efficiency
of information flow is defined and studied in section \ref{e_i_flow}.
Many of the obtained results will be illustrated via an exactly solvable
example of coupled harmonic oscillators; see section \ref{harmo}.
Section \ref{complo} studies the efficiency of heat
flow, while in section \ref{i_flow_t_entropy} we compare the concept of
information flow with the notion of transfer entropy. Our results
are shortly summarized in section \ref{summo}. Some technical questions
are relegated to Appendices. 

\section{Basic concepts: Entropy and mutual information} 
\label{basic}

The purpose of this section is to recall the definition of entropy and
mutual information.

\subsection{Entropy}

How can one quantify the information content of a random variable
$X$ with realizations $x_1,\ldots x_n$ 
and probabilities $p(x_1),\ldots p(x_n)$? This is routinely done by means of the entropy
\BEA
\label{entropy_discrete}
S[X]=-\ssum_{k=1}^n p(x_k)\ln p (x_k).
\EEA
$S[X]$ is well known in physics. Its information-theoretic meaning,
which is based on the law of large numbers, is reminded in Appendix
\ref{info-meaning}. One can arrive at the same form
(\ref{entropy_discrete}) by imposing certain axioms, which are
intuitively expected from the notion of uncertainty or the information
content \cite{ash,strat_IT,cover}. 

For a continuous random variable $X$ the entropy converges to 
\BEA
\label{entropy_cont}
S[X]=-\int \d x \, P(x)\ln P(x)+{\rm\quad additive\,\,\, constant},
\EEA
where $P(x)$ is the probability density of $X$, and where the additive
constant is normally irrelevant, since it cancels out when calculating any
entropy differences. The quantity $e^{-\int \d x \, P(x)\ln P(x)}$
characterizes the [effective] volume of the support of 
$P(x)$ \cite{ash,strat_IT,cover}.

\subsection{Mutual Information}
\label{mumu}

We shall consider the task to predict some present state $Y(t)$ of one
subsystem from the present state $X(t)$ of another subsystem and {\it
vice versa}. For this purpose we introduce two dependent random
variables $X$ and $Y$ with realizations $x_1,\ldots x_n$ and $y_1,\ldots
y_n$ and joint probabilities $\{\,p(x_k,y_l)\,\}_{k,l=1}^n$. 

Assume that we learned a realization $x_l$ of $X$. This allows to
redefine the probabilities of various realizations of $Y$: $p (y_k)\to p
(y_k|x_l)$, where $p (y_k|x_l)$ is the conditional probability. Due to
this redefinition also the entropy $Y$ changes:
$S[Y]\to S[Y|x_l] = -\ssum_{k=1}^n p(y_k|x_l)\ln p (y_k|x_l)$. 
Averaging $S[Y|x_l]$ over $p(x_l)$ we get 
\BEA
S[Y|X] = -\ssum_{k,l=1}^n p(y_k,x_l)\ln p (y_k|x_l).
\nn
\EEA
This conditional entropy characterizes the average resudial entropy of $Y$: $S[Y|X]<S[Y]$
\footnote{Note that generally $S[Y|x_l]<S[Y]$ does not hold, i.e., it is not true that any single
observation reduces the entropy. Such a reduction occurs only in average.}.
If there is a bijective function $f(.)$ such that $Y=f(X)$
($X=f^{-1}(Y)$), then $S[Y|X]=0$. We have $S[Y|X]=S[Y]$ for
independent random variables $X$ and $Y$. 

The mutual information $I[Y:X]$ between $X$ and $Y$ is that part of
entropy of $Y$, which is due to the missing knowledge about $X$.
To define $I[Y:X]$ we subtract the residual entropy
$S[Y|X]$ from the unconditional entropy $S[Y]$:
\BEA
\label{brama}
I[Y:X]&=&S[Y]-S[Y|X]\\
&=&\ssum_{k,l=1}^n p(x_k,y_l)\ln \frac{p(x_k,y_l)}{p_X(x_k)p_Y(y_l)},
\EEA
where $p_X$ and $p_Y$ are the marginal probabilities
$p_X(x_k)=\ssum_{l=1}^n p(x_k,y_l)$ and 
$p_Y(y_l)=\ssum_{k=1}^n p(x_k,y_l)$.

The mutual information is non-negative, $I[Y:X]\geq 0$, symmetric,
$I[Y:X]=I[X:Y]$, and characterizes the entropic response of one variable
to fluctuations of another.  For two bijectively related random
variables $I[X:Y]=S[X]$, and we return to the entropy.  For independent
random quantities $X$ and $Y$, $I[X:Y]=0$. Conversely, $I[X:Y]=0$ imlpies
that $X$ and $Y$ are independent. Thus $I[X:Y]$ is a non-linear correlation function between $X$ and $Y$.

The information-theoretic meaning of the
mutual information $I[X:Y]$ is recalled in Appendix \ref{info-meaning}.
$I[X:Y]$ is related to the information shared via a noisy channel with 
input $X$ and output $Y$, or, alternatively, with input $Y$ and output $X$.

For continuous random variable $X$ and $Y$ with the joint probability
density $P(x,y)$, respectively, the mutual information reads
\BEA
I[X:Y]=\int \d x \, P(x,y)\ln \frac{P(x,y)}{P_X(x)P_Y(y)},
\label{mutu}
\EEA
where additive constants have cancelled after taking the difference in (\ref{brama}).

\section{Model class: Two coupled Brownian particles}
\label{model}

Consider two brownian particles with coordinates $\x=(x_1,x_2)$
interacting with two independent thermal baths at temperatures $T_1$ and
$T_2$, respectively, and subjected to a potential [Hamiltonian] $H(\x)$.
The corresponding time-dependent random variables will be denoted via
$(\,X_1(t), X_2(t)\,)$; their realizations are $(x_1,x_2)$. 

The overdamped limit of the Brownian dynamics is defined by the
following two conditions \cite{risken}: {\it i)} The characteristic
relaxation time of the (real) momenta $m\dot{x}_i$ is much smaller than
the one of the coordinates. This condition is satisfied due to strong
friction and/or small mass \cite{risken}.  {\it ii)} One is interested
in times which are much larger than the relaxation time of the momenta,
but which can be much smaller than [or comparable to] the relaxation
time of the coordinates.  Under these conditions the dynamics of the
system is described by Langevin equations \cite{risken}:
\begin{eqnarray} 
\label{1}
&& 0 = -\partial _i H -\Gamma _i \dot{x}_i
 +\eta _{i}(t),\\ 
&&\langle \eta _{i}(t)\eta_{j}(t^{\prime })\rangle =2\,\Gamma _i \, T_i\, \delta _{ij}\delta (t-t^{\pr })
\quad i,j=1,2,\nonumber
\end{eqnarray}
where $\Gamma _i$ are the damping constants, 
which characterize the coupling of the particles to the respective baths,
$\delta_{ij}$ is the Kronecker symbol, and where 
$\partial_i\equiv \partial /\partial x_i$.
It is assumed that the relaxation time
toward the total equilibrium (where $T_2=T_1$) is much larger than all
considered times; thus for our purposes $T_2$ and $T_1$ are constant
parameters.

Eq.~(\ref{1}) comes from the Newton equation (mass $\times$ acceleration
= conservative force + friction force + random force) upon neglecting
the mass $\times$ acceleration due to strong friction and/or small mass
\cite{risken}. Among many other realizations, Eq.~(\ref{1}) may
physically be realized via two coupled RLC circuits. Then $\Gamma_i=R_i$
corresponds to the resistance of each circuit, $x_i$ is the charge,
while the noise $\eta_i$ refers to the random electromotive
force. The overdamped regime would refer here to small inductance $L_i$
and/or large resistance $R_i$, while the Hamiltonian part $H(x_1,x_2)$
collects separate effects of capacitances $C_1$ and $C_2$, as well as
capacitance-capacitance coupling, e.g.,
$H=\frac{x_1^2}{2C_1}+\frac{x_2^2}{2C_2} +\kappa x_1x_2$ in the harmonic
regime.  This example will be studied in section \ref{harmo}. 

Below we use the following shorthands:
\begin{gather}
\label{barbados}
\partial _t\equiv {\partial}/{\partial t},~
\partial_i\equiv {\partial}/{\partial x_i},~~
\x=(x_1,x_2),~ \d\x=\d x_1\d x_2, \nonumber\\
\y=(y_1,y_2),~ \d\y=\d y_1\d y_2.
\end{gather}

The joint probability distribution $P(x_1,x_2;t)$ satisfies the Fokker-Planck
equation \cite{risken}
\begin{eqnarray}
\label{160}
&&\partial _t P(\x;t) + {\ssum}_{i=1}^2 \partial _i J_i(\x;t) =0, \\
&&\Gamma_i J_i(\x;t)
= -P(\x;t) \,\partial _i H(\x) -T_i
\partial _i P(\x;t),~~
\label{170}
\end{eqnarray}
where $J_1$, $J_2$ are the currents of probability. 
Eq.~(\ref{160}) is supplemented by the standard boundary conditions
\BEA
\label{ozon}
P(x_1,x_2;t)\to 0 ~~ {\rm when}~~ x_1\to\pm\infty~~ {\rm or}~~x_2\to\pm\infty.
\EEA

\subsection{Chapman-Kolmogorov equation.}
\label{ck}

To put our discussion in a more general context, let us recall that 
the process described by (\ref{160}, \ref{170}) is Markovian
and satisfies the Chapman-Kolmogorov equation \cite{risken}:
\BEA
P(\x;t+\tau)=\int \d\y \,P(\x;t+\tau | \y;t) P(\y;t),
\label{chapo}
\EEA
where for $\tau\to 0$ the conditional probability density $P(\x;t+\tau | \y;t)$
is written as \cite{risken}
\begin{gather}
P(\x;t+\tau | \y;t) =
   \delta(x_1-y_1)\delta(x_2-y_2)  ~~~~~~~~~~~~~~~~~~~~\nonumber\\
+ \tau \,\delta(x_2-y_2) \G_1 (x_1|y_1;x_2)+\tau\,\delta(x_1-y_1) \G_2 (x_2|y_2;x_1)\nonumber\\
+{\cal O}(\tau^2), 
\label{veda_1}
\end{gather}
where for $i=1,2$ we have defined
\BEA
\label{veda_2}
&&\G_1(x_1|y_1;x_2)\\
&&\equiv\frac{1}{\Gamma_1}\,
\partial_{1}\left[
\delta(y_1-x_1)\partial_{1}H(\x)+T_1\partial_{1} \delta(y_1-x_1)
\right].\nonumber\\
\label{veda_22}
&&\G_2(x_2|y_2;x_1)\\
&&\equiv\frac{1}{\Gamma_2}\,
\partial_{2}\left[
\delta(y_2-x_2)\partial_{2}H(\x)+T_2\partial_{2} \delta(y_2-x_2)
\right].\nonumber
\EEA
Note from (\ref{veda_1}--\ref{veda_22}) that
\BEA
&&P(x_1;t+\tau|\y;t)\,P(x_2;t+\tau|\y;t)\nonumber\\
&&\,=\,P(x_1,x_2;t+\tau|\y;t)+{\cal O}(\tau^2),
\label{brams}
\EEA
which means that 
the conditional dependence of
$X_1(t+\tau)$ and $X_2(t+\tau)$, given 
$X_1(t)$ and $X_2(t)$ vanishes with second order in $\tau$.

Eqs.~(\ref{160}, \ref{170}) are recovered after substituting
(\ref{veda_1}--\ref{veda_22}) into (\ref{chapo}) and noting
$P(\x;t+\tau)=P(\x;t)+\tau\partial_t P(\x;t)+{\cal O}(\tau^2)$:
\BEA
\pd_t P(\x;t)&=&\int \d y_1 \G_1(x_1|y_1;x_2)\, P(y_1,x_2;t) \nonumber\\
&+& \int \d y_2 \G_2(x_2|y_2;x_1)\, P(x_1,y_2;t). 
\label{litovsk}
\EEA

\section{Information-theoretical definition of information flow}
\label{i_flow}

\subsection{Task} 

We consider the task of predicting gain (or loss) of the future
of subystem $1$ from the present state of subsystem $2$. This task can by
quantified by the information flow $\i_{2\to 1}$, which is defined via the
time-shifted mutual information
\begin{gather}
\label{main_definition}
\i_{2\to 1}(t)=\partial_\tau \,I[\, X_1(t+\tau):X_2(t)\,]\, \,|_{\tau\to +0}~~~~~~~~~~~~~\\
={\rm lim }_{\tau\to +0}\,\frac{1}{\tau}\,\left(I[X_1(t+\tau):X_2(t)]
-I[X_1(t):X_2(t)] \right).\nonumber
\end{gather}
$\i_{1\to 2}(t)$ is defined
analogously with interchanging $1$ and $2$. 

Recall that the mutual information $I[X_1(t+\tau) : X_2(t) ]$ quantifies
(non-linear) statistical dependencies between $X_1(t+\tau)$ and $X_2(t)$, i.e., it
quanti¯es the extent of which the present (at time $t$) of $X_2$ can
predict the future of $X_1$.  Thus a positive $\i_{2\to 1}(t)$ means
that the future of $X_1$ is more predictable for $X_2$ than the present
of $X_1$.  Thus, for $\i_{2\to 1}(t)>0$, $2$ is gaining control over $1$
(or $2$ sends information to $1$) \footnote{For $\i_{2\to 1}(t)>0$, $2$
is like a chief sending orders to its subordinate $1$; the fact that $1$
will behave according to these orders makes its future more predictable
from the vewpoint of the present of $2$.}.  Likewise, $\i_{2\to 1}(t)<0$
means that $2$ is loosing control over $1$, or that $1$ gains autonomy
with respect to $2$. This is the meaning of negative information flow. 

Noting from (\ref{mutu}) that
\BEA
I[\, X_1(t+\tau):X_2(t)\,]=\int \d x_1\d y_2 P_2(y_2;t)\,\times\nonumber\\
P_{1|2}(x_1; t+\tau|y_2;t)\ln \frac{P_{1|2}(x_1; t+\tau|y_2;t)}
{P_{1}(x_1; t+\tau)},\nn
\EEA
we work out $\i_{2\to 1}(t)$ with help of 
(\ref{veda_1}--\ref{veda_22}): 
\begin{gather}
\label{barashek}
\i_{2\to 1} = \int\d \y \,\d x_1 \G_1(x_1|y_1;y_2)P(\y;t) \ln \frac{P_{1|2}(x_1|y_2;t)}
{P_{1}(x_1; t)}.
\end{gather}
Employing now (\ref{160}, \ref{170}) and the boundary
conditions (\ref{ozon}) we get from (\ref{barashek})
\BEA
\label{korkud}
\i_{2\to 1}(t)=\int \d \x
\ln\left[\frac{P_1(x_1;t)}{P(\x;t)}\right]\,\partial_1 J_1(\x;t).
\EEA
Parametrizing the Hamiltonian as
\BEA
H(x_1,x_2)=H_1(x_1)+H_2(x_2)+H_{12}(x_1,x_2),
\label{aka}
\EEA
where $H_{12}(x_1,x_2)$ is the interaction Hamiltonian, we obtain from (\ref{160}, \ref{170}, \ref{korkud})
\BEA
\i_{2\to 1}(t)=\frac{1}{\Gamma_1}\int \d \x\,P(\x;t)\,
\left[\,\partial_1 H_{12}(\x)\right.~~~~~~~~~~~~~~~~\nonumber\\ 
\left. +T_1\partial_1 \ln P(\x;t )\,\right]
\partial_1 \ln \frac{P_1(x_1;t)}{P(\x;t)}.
\label{choban}
\EEA

The time-shifted mutual information was employed for quantifying information flow 
in reaction-diffusion systems \cite{vastano}, neuronal ensembles \cite{neuro_vastano},
coupled map lattices \cite{ho_vastano}, 
and ecological dynamics \cite{nichols_vastano}. 

\comment{
Before continuing on the physical meaning of information flow in section
\ref{basso}, let us mention that the definition (\ref{main_definition})
can be recovered from an operational approach suggested in \cite{majda};
see also \cite{liang_kleeman_prl,liang_kleeman_pd} for related works.
This fact is demonstrated in Appendix \ref{opera}. The main point of
this approach is that the information flow $\i_{2\to 1}$ is introduced
via a direct physical action, namely the freezing of the dynamics of the
second particle. 
Thus, $\i_{2\to 1}(t)$ is that part of the entropy change of
$X_1$ (between $t$ and $t+\tau$), which exists due to fluctuations
of $X_2(t)$; see section \ref{mumu}.  }

\subsection{Basic features of information flow}
\label{basso}

{\bf 1.} 
As deduced from (\ref{korkud}), the information flow is generally not symmetric
\BEA
\i_{2\to 1}(t)\not=\i_{1\to 2}(t),\nn
\EEA
but the symmetrized information flow (\ref{main_definition})
is equal to the rate of mutual information
\BEA
\label{mangust}
\label{brat}
&&\i_{2\to 1}(t)+\i_{1\to 2}(t) = \frac{\d}{\d t}\, I[X_1(t):X_2(t)].
\label{kora}
\EEA

While the mutual information is symmetric with respect to inter-changing
its arguments $I[X_1(t):X_2(t)]=I[X_2(t):X_1(t)]$ and it quantifies
correlations between two random variables, the information flow is
capable of distinguishing the source versus the recipient of
information: $\i_{2\to 1}(t)>0$ means that 2 is the source of
information, and 1 is its recipient. 

For $\i_{2\to 1}(t)>0$ {\it and} $\i_{1\to 2}(t)>0$ we have a {\it
feedback regime}, where 1 and 2 are both sources and recipients of
information.  Now the interaction between 1 and 2 builts up the mutual
information $I[X_1(t):X_2(t)]$; see (\ref{mangust}).  In contrast, for
$\i_{2\to 1}(t)<0$ and $\i_{1\to 2}(t)<0$ both particles are detached
from each other, and the mutual information naturally decays. 

For $\i_{2\to 1}(t)>0$ and $\i_{1\to 2}(t)<0$ we have one-way flow of
information: 2 is source and 1 is recipient; likewise, for $\i_{2\to
1}(t)<0$ and $\i_{1\to 2}(t)>0$, 1 is source and 2 is recipient. A
particular, but important case of this situation is when the mutual
information is conserved: $\frac{\d }{\d t} I[X_1(t):X_2(t)] =0$. This
is realized in a stationary case; see below.  Now due to $\i_{2\to
1}(t)+\i_{1\to 2}(t)=0$ the information behaves as a conserved resource
(e.g., as energy): the amount of information lost by 2 is received by 1,
and {\it vice versa}. Another example of one-way fow of information is
$\i_{2\to 1}(t)>0$ and $\i_{1\to 2}(t)\simeq 0$.

{\bf 2.} The information flow $\i_{2\to 1}$ can be represented as
[see (\ref{brama}, \ref{main_definition})]
\BEA
\i_{2\to 1} = \frac{\d S[X_1(t)]}{\d t} -\pd_\tau S[X_1(t+\tau)\, |\, X_2(t)]|_{\tau\to 0}.
\label{sw}
\EEA
The first term in the RHS of (\ref{sw}) is the change of the marginal
entropy of $X_1$, while the second term is the change of the conditional
entropy of $X_1$ with $X_2$ being {\it frozen} to the value $X_2(t)$. In
other words, $\i_{2\to 1}(t)$ is that part of the entropy change of
$X_1$ (between $t$ and $t+\tau$), which exists due to fluctuations of
$X_2(t)$; see section \ref{mumu}. This way of looking at the information
flow is close to that suggested in \cite{majda}; see also
\cite{liang_kleeman_prl,liang_kleeman_pd} for related works. Appendix
\ref{opera} studies in more detail the operational meaning of the freezing
operation.

{\bf 3.} Eq.~(\ref{choban}) implies the information flow $\i_{2\to 1}(t)$ can be 
divided in two components: a force-driven part $\i_{2\to 1}^{\rm F}(t)$ and bath-driven (or fluctuation-driven) part 
$\i_{2\to 1}^{\rm B}(t)$
\BEA
&&\i_{2\to 1}(t)=\i_{2\to 1}^{\rm F}(t)+\i_{2\to 1}^{\rm B}(t), \nonumber\\
&&\i_{2\to 1}^{\rm F}(t)\equiv\frac{1}{\Gamma_1}\int \d \x\,P\,
[\partial_1 H_{12}]\,\,
\partial_1 \ln \frac{P_1}{P},\\
\label{hoviv}
&&\i_{2\to 1}^{\rm B}(t)= -\frac{T_1}{\Gamma_1}\int \d \x\,P_1\,P_{2|1}\,[\pd_1 \ln P_{2|1}]^2,
\label{brandt}
\EEA
where for simplicity we omitted all integration variables. 
The force-driven part $\i_{2\to 1}^{\rm F}(t)$ nullifies together with
the force $-\partial_1 H_{12}$ acting from the second particle on the
first particle. 
The fluctuation-driven part $\i_{2\to 1}^{\rm B}(t)$
nullifies together with the bath temperature $T_1$ (which means that the
random force acting from the first bath is zero). 

Note that although $\i_{2\to 1}^{\rm F}(t)$ is defined
via the interaction Hamiltonian $H_{12}$, it does not suffer from the
known ambiguity related to the definition of $H_{12}$. That is
redefining $H_{12}(\x)$ via $H_{12}(x)\to H_{12}(x)+f_1(x_1)+f_2(x_2)$,
where $f_1(x_1)$ and $f_2(x_2)$ are arbitrary functions will not alter
$\i_{2\to 1}^{\rm F}(t)$ (and will not alter $\i_{2\to 1}(t)$, of
course). Thus, changes in separate Hamiltonians $H_1(x_1)$
and $H_2(x_2)$ that do not alter the probabilities (e.g., sudden
changes) do not influence $\i_{2\to 1}^{\rm F}(t)$. In contrast, sudden
changes of the interaction Hamiltonian will, in general, contribute to
$\i_{2\to 1}^{\rm F}(t)$. 

The bath-driven contribution $\i_{2\to 1}^{\rm B}(t)$ into the
information flow is negative, which means that for $2$ to be a source of
information, i.e., for $\i_{2\to 1}(t)>0$, the force-driven part
$\i_{2\to 1}^{\rm F}(t)$ should be sufficiently positive. In short,
there is no transfer of information without force. 

{\bf 4.} It is seen from (\ref{korkud}) [or from (\ref{choban})] that if
the variables $X_1$ and $X_2$ are independent, i.e.,
$P(\x;t)=P_1(x_1;t)P_2(x_2;t)$ the information flow $\i_{2\to 1}$
nullifies:
\BEA
\label{desht}
\i_{2\to 1}(t)=\i_{1\to 2}(t) =0~~ {\rm for}~~ P=P_1P_2.
\EEA
In fact, both $\i_{2\to 1}^{\rm F}(t)$ and $\i_{2\to 1}^{\rm B}(t)$ nullify for
$P=P_1P_2$. 

If the two particles were interacting at times $t<t_{\rm switch}$, but
the interaction Hamiltonian $H_{12}$ [see (\ref{aka})] is switched off
at $t=t_{\rm switch}$ the information flow $\i_{2\to 1}=\i_{2\to 1}^{\rm
B}(t)$ will be in general different from zero for times $t_{\rm
relax}+t_{\rm switch}>t>t_{\rm switch}$|where $t_{\rm relax}$ is the
relaxation time| since at these times the common probability will be
still non-factorized, $P\not=P_1P_2$. However, since now $\i_{2\to
1}(t) = \i^{\rm B}_{2\to 1}(t)<0$ the particles can only decorrelate from
each other: neither of them can be a source of information for another. 

We note that $\i^{\rm B}_{2\to 1}(t)=-\frac{T_1}{\Gamma_1}\int \d x_1 P_1(x_1){\cal F}_1(x_1)$ 
in (\ref{brandt}) is proportional to the average Fisher
information ${\cal F}_1(x_1)$ \cite{cover}:
\BEA
{\cal F}_1(x_1)\equiv\int \d x_2\, P_{2|1}(x_2|x_1)\,[\pd_1 \ln P_{2|1}(x_2|x_1)]^2.
\nn
\EEA
$1/{\cal F}_1(x_1)$ is the minimal variance that can be reached during any
unbiased estimation of $x_1$ from observing $x_2$ \cite{cover}.
In the sense of estimation theory ${\cal F}_1(x_1)$ quantifies the information
about $x_1$ contained in $x_2$, since ${\cal F}_1(x_1)$ is larger for those
distributions $P_{2|1}(x_2|x_1)$ whose support is concetrated at
$x_2\simeq x_1$ \,\footnote{For various applications of the Fisher information
in the physics of Brownian motion see \cite{garba}.}. We conclude that once the
interaction between $1$ and $2$ is switched off, $1$ gets detached (diffuses away)
from $2$ by the rate equal to the diffusion constant $\frac{T_1}{\Gamma_1}$ times
the average Fisher information about $x_1$ contained in $x_2$.

\begin{figure}
\centerline{\includegraphics[scale=0.80]{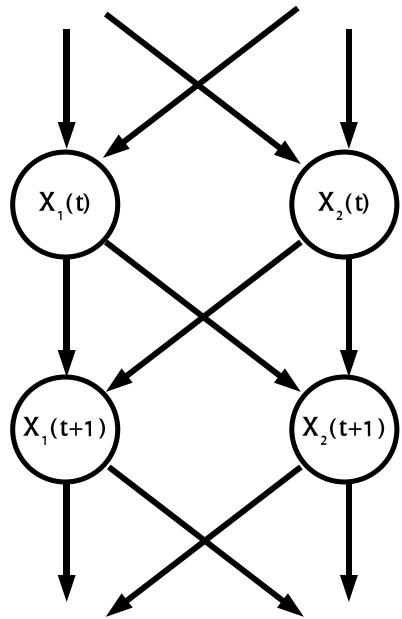}}
\caption{\label{Gr}{\small Time-discretized version of the joint dynamical evolution.}}
\end{figure}

{\bf 5.} To visualize information flow, we time-discretize the
two-particle system and consider the two time series $X_1(t), X_2(t)$
with $t=1,2,\ldots$. We model the system dynamics by a first order
Markov process, as being the discrete analogue of a first order
differential equation.  It is described by the graphical model depicted
in Fig.~\ref{Gr}. 

The graph is indeed meant in the sense of a causal structure
\cite{Pearl}, where an arrow indicates direct causal influence. In
particular, the arrows on Fig.~\ref{Gr} mean that $X_1(t+1)$ and
$X_2(t+1)$ become independent after conditiong over $(X_1(t),\,
X_2(t)\,)$; see (\ref{brams}) in this context.  Assume, for the moment,
that the arrrow between $X_2(t)$ and $X_1(t+1)$ were missing, e.g.,
$X_2(t)$ is simply a passive obervation of $X_1(t)$.  Then 
$X_2(t)$ and $X_1(t+1)$ are conditionally independent given $X_1(t)$. Now
the data
processing inequality \cite{cover} imposes negativity of the
(time-discretized) information flow
\begin{equation}\label{DataProcess}
I[X_1(t+1):X_2(t)]- I[X_1(t):X_2(t)]\leq 0\,,
\end{equation}
because $X_1(t+1)$ is obtained from $X_1(t)$ by a stochastic map (which
can never increase the information about $X_2(t)$ without having access
to this quantity). Expression~(\ref{DataProcess}) can, however, be
properly below zero if the bath dissipates some of the information.  If
(\ref{DataProcess}) is positive there must be an arrow from $X_2(t)$ to
$X_1(t+1)$, i.e., there is information flowing from $X_2$ to $X_1$.
However, the presence of such an arrow does not guarantee positivity of
(\ref{DataProcess}) because the amount of information dissipated by the
bath can exceed the one provided by the arrow. We can nevertheless interpret
$I[X_1(t+1):X_2(t)]- I[X_1(t):X_2(t)]$
as the information in the sense of a net effect.

\section{Thermodynamic aspects of information flow}
\label{i_flow_thermo}

\subsection{Information flow nullifies in equilibrium}

So far we have discussed the formal definition of information flow $\i_{2\to
1}$. This definition relates $\i_{2\to 1}$ to the concept of mutual
information. It is however expected that there should be a more physical
way of understanding $\i_{2\to 1}$, since the concepts of entropy and
entropy flow are defined and discussed in thermodynamics of
non-equilibrium systems \cite{hb,berg,Meixner,Onsager}. We now turn to this aspect. 

Recall that the Fokker-Planck equation (\ref{160}, \ref{170}) describes
a system interacting with two thermal baths at different temperatures
$T_1$ and $T_2$.  If $T_1=T_2=T$ then the two-particle system relaxes
with time to the Gibbs distribution with the common temperature $T$
\cite{risken}:
\BEA
\label{gibbs}
P_{\rm eq}(\x)=\frac{1}{Z}\, e^{-\beta H(\x)}, 
\qquad Z=\int\d \x  \, e^{-\beta H(\x)}.
\EEA

An important feature of the {\it equilibrium} probability distribution (\ref{gibbs}) is that the
currents of probability (\ref{170}) do not depend on time and explicitly nullify in that state 
\BEA
\label{jabar}
J_1(\x)=J_2(\x)=0.
\EEA
This {\it detailed balance} feature| which is both necessary and
sufficient for equilibrium| indicates that in the equilibrium state
there is no transfer of any physical quantity, e.g., there is no
transfer of energy (heat). 

Eq.~(\ref{korkud}) implies that the same holds for $\i_{2\to 1}$: 
in the equilibrium state there is no information flow,
\BEA
\i_{2\to 1}=\i_{1\to 2}=0.
\nn
\EEA
In other words, 
the transfer of information should always be connected with a certain
non-equilibrium situation. One may distinguish several types of such situations: 

{\it i)}
Non-stationary (transient) states, where the joint distribution $P(x_1,x_2;t)$ is
time-dependent. 

{\it ii)} Stationary state, but non-equilibrium states realized for $T_1\not
=T_2$. Now the probability currents $J_1$ and $J_2$ are not zero (only
$\partial_1 J_1+\partial_2 J_2=0$ holds), and so are the information
transfer rates $\i_{2\to 1}$ and $\i_{1\to 2}$.  However, since the
state is stationary, their sum nullifies due to (\ref{kora})
\BEA
\label{samson}
\i_{2\to 1}+\i_{1\to 2}=0.
\EEA
This on indicates on one-way flow of information.
Below we clarify how this flow relates to the temperature difference. 

{\it iii)} Non-equilibrium states can be maintained externally by
time-varying conservative forces, or, alternatively, by time-independent
non-conservative forces accompanied by cyclic boundary conditions. 
Here we do not consider this type of non-equilibrium. 

\subsection{Entropy production and heat dissipation.}

A non-equilibrium state will show a tendency towards equilibrium,
or, as expressed by the second law, by the entropy production of the
overall system (in our case the brownian particles plus their
thermal baths). Since for the model (\ref{1}, \ref{160}, \ref{170}) the
baths are in equilibrium and the cause of non-equilibrium is related to
the brownian particles, the overall entropy production can be expressed
via the variables pertaining to the brownian particles \cite{hb,berg,Meixner}. 
Recall (\ref{litovsk}) and define for $i=1,2$ 
\BEA
\label{u1}
\frac{\dbarrm _i Q}{\d t}&=&-\int \d \x\, H(\x) \,
\partial _i J_i(\x;t),\\ 
\label{u2}
\frac{\dbarrm _i S}{\d t}&=&\int \d \x \,[\,\ln P(\x;t) \,]\,
\partial _i J_i(\x;t),
\label{urm_1}
\EEA
and note that these quantities satisfy
\BEA
\label{khajam}
\ell_i&=& \frac{\dbarrm _i S}{\d t}-\beta_i \frac{\dbarrm _i Q}{\d t}\\
&=&\beta_i\Gamma_i\int\d \x\,\frac{J_i^2(\x;t) }{P(\x;t) }\geq 0,
\label{omar}
\EEA
where $\ell_1=\ell_2=0$ holds in the 
equilibrium only, where $J_1=J_2=0$; 
see (\ref{jabar}).

Our discussion in section \ref{ck} implies that
$\frac{\dbarrm _i Q}{\d t}$ is the change of energy of the two-particle
system due to the dynamics of $x_i$.  Since in the overdamped regime
this dynamics is driven by the bath at temperature $T_i$, we see that
$\frac{\dbarrm _i Q}{\d t}$ is the heat received by the two-particle
system from the bath at temperature $T_i$ \cite{Meixner,Onsager}. Thus the 
energy (heat) recieved by the bath is $-\frac{\dbarrm _i Q}{\d t}$.
Likewise, $\frac{\dbarrm _i S}{\d t}$ is the change of entropy of the
two-particle system due to the dynamics of $x_i$.  Then (\ref{omar})
is the local version of the Clausius inequality \cite{Meixner,Onsager},
which implies that $\ell_i$ is the [local] entropy produced per time due to the
interaction with the bath at temperature $T_i$, while
\BEA
\label{total}
\ell=
\ell_1+\ell_2 = \frac{\d S}{\d t}-\beta_1\frac{\dbarrm _1 Q}{\d t}-\beta_2\frac{\dbarrm _2 Q}{\d t}
\EEA
is the total entropy produced per unit of time in the overall system:
two {\it equilibrium} baths plus two brownian particles \cite{hb,risken,Meixner,Onsager}. 

\comment{\footnote{We should stress again that the above interpretation of $\frac{\dbarrm
_i Q}{\d t}$ and $\frac{\dbarrm _i S}{\d t}$ is grounded on the specific
form (\ref{veda_1}) of the two-particle short-time transition
probability; see our discussion in the end of section \ref{ck}. These
interpretations will in general not hold without this explicit form. }}

Noting our discussion in Appendix \ref{coarse-grained} one can see that
$\frac{\ell_i}{\beta_i\Gamma_i}$ is equal to the space-average $\int\d
\x P(\x;t)v^2_i(\x;t)$, where $v_i(\x;t)$ is the coarse-grained
velocity; see (\ref{katu}, \ref{shun})
\footnote{\label{aram_2}Note from (\ref{omar}) that for $T_1=T_2$ 
|where the non-stationarity is the sole cause of non-equilibrium|
the heat dissipation, or the total entropy production $\ell$ times $T_1=T_2=T$, is 
equal to the negative rate of free energy \cite{risken}:
$-T\ell=\frac{\d F}{\d t}
\equiv\frac{\d }{\d t}
\int \d \x \,[H(\x) +T\ln P(\x;t) \,]\,P(\x;t)$.
Note that the free energy here is defined with respect to the bath
temperature $T$ and that $P(\x;t)$ is an arbitrary non-equilibrium
distribution that relaxes to the equilibrium: $P_{\rm eq}(\x)\propto
\exp\left[-\beta H(\x)\right]$. The difference $F-F_{\rm eq}$
between the non-equilibrium and equilibrium free energy is the maximal
work that can be extracted from the non-equilibrium system in contact
with the thermal bath \cite{ll}. 
}. Similarly, integrating
(\ref{u1}) by parts one notes that the $-\frac{\dbarrm _i Q}{\d t}$
amounts to the space-averaged work done by the potential force
$-\frac{\dbarrm _i Q}{\d t}=\int \d \x\, P(\x;t)v_i(\x;t)\partial _i
H(\x)$ \footnote{A similar argument to the physical meaning of $\frac{\dbarrm _i Q}{\d t}$
can be given via stochastic energetics \cite{sekimoto}.}. 

Note that in general $\ell_i$ (and thus $\ell$) are non-zero for the
non-equilibrium stationary state realized for $T_1\not= T_2$, since then
$J_i(\x)\not =0$. This is consistent with the interpretation of these
states as ``metastable'' non-equilibrium states, where some heat flows
between the two baths, but the temperatures $T_i$ do not change in time
due to the macroscopic size of the baths (these temperatures would
change for very large times, which are, however, beyond the time-scales
considered here.)

Likewise, $T_i\ell_i$ is the heat dissipated per time due to the
interaction with the bath at temperature $T_i$, while
$T_1\ell_1+T_2\ell_2$ is the total dissipated energy (heat) per time
\cite{Onsager}.

\subsection{
Equivalence
between flow of mutual information and flow of entropy
}

The picture that emerges from the above consideration is as follows. The
entropy is produced with the rate $\ell_1$ somewhere at the interface
between the first brownian particle and its bath. Eq.~(\ref{omar})
expresses the fact that $\frac{\dbarrm _1 S}{\d t}$ is the rate by which
a part of the produced entropy flows into the two-particle system. The
rest of the produced entropy goes to the bath with the rate $-\beta_1
\frac{\dbarrm _1 Q}{\d t}$. This is consistent with
the conservation of energy and the fact that the bath itself is in
equilibrium; then $-\frac{\dbarrm _1 Q}{\d t}$ is the rate by which
heat goes to the bath, and after dividing by $T_1$|since the
bath is in equilibrium|this becomes the rate with which the bath
receives entropy. 

The corresponding argumentation can be repeated for the second particle and its thermal bath.

The  thermodynamic definition of entropy flow eventually  reads
\BEA
\label{luter}
\i_{2\to 1} =\frac{\d S_1}{\d t}-\frac{\dbarrm _1 S}{\d t}.
\EEA
Once $\frac{\dbarrm _1 S}{\d t}$
is the entropy entering into the 
two-particle system via the first particle, then subtracting $\frac{\dbarrm _1 S}{\d t}$
from the entropy rate $\frac{\d S_1}{\d t}$ of the first particle itself, we 
get the entropy flow from the second particle to the first one. 

It should be now clear that this definition of entropy flow is just equivalent to the
definition of information flow (\ref{main_definition}).
Eq.~(\ref{luter}) can also be written as 
\BEA
\label{muter}
\i_{2\to 1} =\frac{\dbarrm _1 S_1}{\d t}+\frac{\dbarrm _1 S_2}{\d t}-\frac{\dbarrm _1 S}{\d t}
=\frac{\dbarrm _1 S_1}{\d t}-\frac{\dbarrm _1 S}{\d t},
\EEA
making clear again that $\i_{2\to 1}$ is the change of mutual information due to dynamics of the
first particle.

\section{Efficiency of information flow}
\label{e_i_flow}

Once it is realized that non-zero information transfer is possible
only out of equilibrium, the existence of such a transfer is related to
entropy production. This is an entropic cost of the information
transfer. One can define a dimensionless
ratio:
\BEA
\label{dalila}
\eta_{2\to 1}=\frac{\i_{2\to 1}}{\ell},
\EEA
which is the desired output (= information transfer) over the
irreversibility cost (= total entropy production). This quantity characterizes
the efficiency of information transfer. A large $\eta$ is desirable,
since it gives larger information transfer rate at
lesser cost. 

Note that $\eta_{2\to 1}$ is more similar to the coefficient of
preformance of thermal refrigerators|which is also defined as the useful
output (heat extracted from a colder body) to the cost (work)|than to
the efficiency of heat engines. The latter is defined as the useful
output over the resource entered into the engine. We shall
still call $\eta_{2\to 1}$ efficiency, but this distinction is to be
kept in mind.

Below in several different situations we shall establish upper bounds on
$\eta$. These determine the irreversibility (entropy) cost of information
transfer. The usage of the efficiency for informational processes was
advocated in \cite{pop}. 

\subsection{Stationary case}

In the stationary two-temperature scenario the joint probability
$P(x_1,x_2)$ and the probability currents $J_1(x_1,x_2)$ and
$J_2(x_1,x_2)$ do not depend on time. Thus many observables|e.g., the
average energy of the two Brownian particles, their entropy, entropies
of separate particles| do not depend on time either. 

Employing (\ref{total}), 
$\i_{2\to 1} =-\frac{\dbarrm _1 S}{\d t}=\frac{\dbarrm _2 S}{\d t}$ 
and (\ref{khajam}) we get
\begin{gather}
\ell=(\beta_2-\beta_1) \frac{\dbarrm _1 Q}{\d t},~~~~
T_2\,\i_{2\to 1}+\frac{\ell}{\beta_2-\beta_1}=T_2\ell_2\geq 0, \nn
\\
\label{hod1}
\i_{2\to 1}=\frac{T_1 \ell}{T_2-T_1}+\ell_2.
\end{gather}
After interchanging the indices $1$ and $2$ we get the analogous equation for 
$\i_{1\to 2}$. Recalling that (\ref{samson}) holds in the stationary state, we 
get 
\BEA
\label{hod2}
\i_{2\to 1}=\frac{T_2 \ell}{T_2-T_1}-\ell_1.
\EEA
First of all, (\ref{hod1}) implies that $\i_{2\to 1}\geq 0$ for $T_2\geq
T_1$, which means that in the stationary two-temperature scenario the
information (together with heat) flows from higher to
lower temperature. 

For the efficiency (\ref{dalila}) we get from (\ref{hod1}, \ref{hod2}):
\BEA
\label{bala1}
\eta_{2\to 1}&=&\frac{T_1}{T_2-T_1}+\frac{\ell_2}{\ell_1+\ell_2}\\
             &=&\frac{T_2}{T_2-T_1}-\frac{\ell_1}{\ell_1+\ell_2}.
\label{bala2}
\EEA
This then implies [since $\ell_1\geq 0$, $\ell_2\geq 0$]
\BEA
\frac{T_1}{T_2-T_1}\leq \eta_{2\to 1} \leq \frac{T_2}{T_2-T_1}.
\label{wolf}
\EEA
For $T_2>T_1$, $\eta_{2\to 1}$ is positive and is bounded from above by
$\frac{T_2}{T_2-T_1}$, which means that in the stationary state the
ratio of the information transfer (rate) over the entropy production
(rate) is bounded. 

It is important to note from (\ref{wolf}) that for $T_2$ approaching
$T_1$ from above, $T_2\to T_1$, the efficiency of information transfer
tends to plus infinity, since in this reversible limit the entropy
production, which should be always positive, naturally scales as
$(T_1-T_2)^2$, while the information flow scales as $T_2-T_1$. Thus very
slow flow of information can be accompanied by very little entropy
production such that the efficiency becomes very large. A similar
argument led some authors to conclude that there is no fundamental cost
for the information transfer at all
\cite{landauer_communication,toffoli,bennett_brownian}. Our analysis
makes clear that this interpretation would be misleading. It is more
appropriate to to say that in the reversible limit the thermodynamic
cost, while becoming less restrictive [since the efficiency can be very
large], is certainly still there, because the difference $T_2-T_1$ is,
after all, always finite; otherwise the very information flow would
vanish. 

It remains to stress that the relations (\ref{hod1}--\ref{wolf}) are
general and do not depend on the details of the considered Brownian
system. They will hold for any bi-partite Markovian system which
satisfies to master equation (\ref{litovsk}) and local formulations of
the second law (\ref{khajam}, \ref{omar}). 

\subsection{Reachability of the upper efficiency bound}
\label{richting}

In the remaining part of this section we show that the upper bound (\ref{wolf})
for the efficiency of information flow is reached in a certain class of Brownian
systems satisfying time-scale separation.

\subsubsection{The stationary probability in the adiabatic case.}
\label{aadiaa}

For $T_1=T_2$, the stationary probability distribution $P(\x)$ of the
two-particle system (\ref{160}, \ref{170}) is Gibbsian: $P(\x)\propto
e^{-\beta_1 H(\x)}$. For non-equal temperatures $T_1\not= T_2$ a general
expression for this stationary probability can be derived in the
adiabatic situation, where $x_2$ changes in time much slower than
$x_1$ \cite{LandauerWoo,Onsager,Ritter,Coolen}. This is ensured by 
\BEA
\label{astrakhan}
\varepsilon 
\equiv \Gamma_1/\Gamma _2\ll 1. 
\EEA
Below we present a
heuristic derivation of the stationary distribution $P(\x)$ in the order
$\varepsilon^0$ \cite{LandauerWoo,Onsager,Ritter,Coolen}; a more systematic
presentation, as well as higher-order corrections in $\varepsilon $ are discussed
in \cite{Onsager}. 

On the time scales relevant for $x_1$, the variable $x_2$ is fixed,
and the conditional probability $P_{1|2}(x_1|x_2)$ is Gibbsian [$\beta_1=1/T_1$]:
\BEA
\label{bora77}
P_{1|2}(x_1| x_2)=\frac{e^{-\beta _1 H(\x)}}{Z(x_2)}, ~~
Z(x_2)=\int \d x_1\, e^{-\beta _1 H(\x)}.
\EEA
The stationary probability
$P_2(x_2)$ for the slow variable is found by noting that on the times relevant for $x_2$, $x_1 $ is
already in the conditional steady state. Thus the force $ \partial_2
H(x_1,x_2)$ acting on $x_2$ can be averaged over $P_{1|2}(x_2|x_1)$
\BEA
\int\d x_1  \, \partial_2
H(\x) P_{1|2}(x_1|x_2)=\partial_2 F(x_2), 
\nn
\EEA
where $F(x_2)=-T_1\ln Z(x_2)$ is the conditional free energy.  Once the averaged
dynamics of $x_2$ is governed by the effective potential energy
$F(x_2)$, the stationary $P_2(x_2)$ is Gibbsian at the inverse temperature
$\beta_2=1/T_2$:
\BEA
P_2(x_2)=\frac{e^{-\beta_2 F(x_2)}}{\CZ  },\quad \CZ=\int\d x_2 \, e^{-\beta_2 F(x_2)}.
\nn\EEA
Thus the joint stationary probability is 
\begin{gather}
\label{osh}
P(\x)=P_2(x_2)P_{1|2}(x_1|x_2)=\frac{1}{\CZ}\,e^{(\beta_1-\beta_2)F(x_2)-\beta_1 H(\x)}.
\end{gather}

For further calculations we shall need 
the stationary current $J_2(\x)$, which reads from (\ref{170}, \ref{aka}, \ref{osh})
\BEA
\label{190}
&&
J_2(\x)=\frac{T_1-T_2}{T_1\Gamma_2}P(\x)\, \phi(\x),   \\
&&\phi_{1\to 2} (\x)\equiv 
-\partial _2 H_{12}(\x) +
\int \d y P_{1|2}(y |x_2)\partial _2 H_{12} (y,x_2),\nn\\
&&
\label{191}
\EEA
where $\phi_{1\to 2}(\x)$ is the force acting on $2$ from $1$ minus its average over
the fast $x_1$. Both $J_1(\x)$ and $J_2(\x)$ are of the same order ${\cal
O}(\frac{\varepsilon }{\Gamma _1})= {\cal O}(\frac{1}{\Gamma _2})$. 

Let us describe how to obtain $J_1(\x)$ (we shall need below at least an order of magnitude estimate of $J_1$). Upon
substituting (\ref{osh}) into (\ref{170}) we find that the current
$J_1(\x)$ nullifies. This means that $J_1(\x)$ has to be searched for to
first order in $\varepsilon =\Gamma_1/\Gamma _2$. Assuming for the corrected probability
$\widetilde{P}(\x)=P(\x)[1-\varepsilon 
{\cal A}(\x)]+{\cal O}(\varepsilon ^2)$ we get 
\BEA
\label{riga}
J_1(\x)=\varepsilon \frac{T_1}{\Gamma _1}P(\x)\partial_1 {\cal A}(\x),
\EEA 
where ${\cal A}(\x)$ does not depend on $\varepsilon $ and is to be found from the
stationarity equation $\partial_1 J_1+\partial_2 J_2=0$. 
Concrete expressions for ${\cal A}(\x)$ are presented in \cite{Onsager}.

\subsubsection{Entropy production and heat dissipation in the adiabatic stationary case.}

Using (\ref{omar}) together with (\ref{190}) and (\ref{riga}) 
we get for the partial entropy productions:
\BEA
\label{soso0}
&&\ell_1 = \varepsilon \,\frac{T_1}{\Gamma_2}\int\d \x\, [\partial_1{\cal A}(\x)]^2\, P(\x), \\
\label{soso1}
&&\ell_2 = \frac{(T_1-T_2)^2}{T_2T_1^2\Gamma_2}\int\d \x \,\phi_{1\to 2}^2(\x)\,P(\x).
\EEA

Note that $\ell_1$ contains an additional small factor $\varepsilon $ as
compared to $\ell_2$.  This is natural, since the fast system $x_1$
is in a local thermal
equilibrium. Thus in the considered order $\varepsilon^0$ the overall entropy
production $\ell$ is dominated by the entropy production of the slow
sub-system:
\begin{equation}
\label{soso2}
\ell_1=0, \qquad 
\ell=\ell_2.
\end{equation}
Thus the heat dissipation is simply $T_2\ell=T_2\ell_2$.
The physical meaning of $\ell_1=0$ is that the fast system
does not produce entropy [and does not dissipate heat], since in its
fast characteristic times it sees the slow variable as a frozen [not
fluctuating] field. Thus the fast system remains in local equilibrium. 

\subsubsection{Efficiency of information transfer in the adiabatic stationary case.}

Employing (\ref{soso2}) we can immediately see from (\ref{bala2}) that the efficiency 
in the considered order of $\varepsilon$ reads:
\BEA
\eta_{2\to 1}=\frac{T_2}{T_2-T_1}.
\label{fox}
\EEA
Thus, the efficiency reaches its maximal value when the hotter system is the slowest one.

Recall that in the
adiabatic limit the information flow is of order ${\cal
O}(1/\Gamma_2)$, i.e., it is small on the characteristic time scale of the
fast particle, but sizable on the characteristic time of the slow particle. 
In more detail, using (\ref{osh}, \ref{190}) 
we get for the information flow $\i_{2\to 1}$
from the slow to the fast particle:
\BEA
\label{soso3}
\i_{2\to 1} = \frac{T_2-T_1}{T_1^2\Gamma_2}\int\d \x \, \phi_{1\to 2}^2(\x)P(\x).
\EEA
The reachability of the upper bound for the efficiency $\eta_{2\to 1}$
appears to resemble the reachability of the optimal Carnot efficiency
for heat engines. However, for heat engines the Carnot efficiency is
normally reached for processes that are much slower than any internal
characterisitc time of the engine working medium. In contrast, the
information flow in (\ref{fox}) is sizable on the time-scale $\Gamma_2$
(which is one of the internal time-scales).

\section{Exactly solvable model: Two coupled harmonic oscillators}
\label{harmo}

We examplify the obtained results by the exactly
solvable model of two coupled harmonic oscillators. We shall also employ
this model to check whether the upper bound (\ref{wolf}) may hold in the
non-stationary situation. 

The Hamiltonian [or potential energy] is given as
\BEA
\label{hamo}
H(\x)=\frac{a_1}{2}x_1^2+\frac{a_2}{2}x_2^2+b x_1x_2,
\EEA
where the constants $a_1$, $a_2$ and $b$ have to satisfy
\BEA
a_1>0, \quad a_2>0, \quad a_1a_2>b^2,
\label{balasan}
\EEA
for the Hamiltonian to be positively defined. 

Let us assume that the probability of the two-particle
system is Gaussian [this holds for the stationary probability as seen below]
\BEA
-\ln P(\x)=\frac{A_1}{2}x_1^2+\frac{A_2}{2}x_2^2+B x_1x_2+{\rm constant},
\label{yara}
\EEA
Here we did not specify the irrelevant normalization constant, and 
the constants $A_1$, $A_2$ and $B$ read
\BEA
\left(\begin{array}{rr}
A_1 & B~ \\
B~  & A_2\\
\end{array}\right)^{-1}
=\left(\begin{array}{rr}
\langle x_1^2\rangle~ & \langle x_1 x_2\rangle \\
\langle x_1 x_2\rangle  & \langle x_2^2\rangle~\\
\end{array}\right)\equiv \left(\begin{array}{rr}
\sigma_{11} & \sigma_{12} \\
\sigma_{12}  & \sigma_{22}\\
\end{array}\right),\nonumber\\
\label{hajro}
\EEA
where $\langle\ldots\rangle$ is the average over the probability distribution (\ref{yara}).
All the involved parameters ($A_1$, $A_2$ and $B$) can be in general time-dependent.
Note that in (\ref{yara}) we assumed $\langle x_1\rangle=\langle x_2\rangle=0$.

For the Hamiltonian (\ref{hamo}) and the probability (\ref{yara}) the information transfer reads from 
(\ref{choban}):
\BEA
\label{kob1}
\i_{2\to 1}=\frac{\sigma_{12}(T_1B-b)}{\Gamma_1\sigma_{11}}.
\EEA
Likewise, we get for the local entropy production (\ref{omar}):
\BEA
\ell_1&=&\frac{T_1}{\Gamma_1}\, [\,\sigma_{11}(\beta_1 a_1 -A_1)^2+\sigma_{22}(\beta_1 b -B)^2\nonumber\\
      &+&      2\sigma_{12}(\beta_1 a_1 -A_1)(\beta_1 b -B) \,],
\label{kob2}
\EEA
where the information transfer $\i_{1\to 2}$ and the entropy production
$\ell_2$ are obtained from (\ref{kob1}) and (\ref{kob2}), respectively,
by interchanging the indices $1$ and $2$. 

\subsection{Stationary case.}

Provided that the stability conditions (\ref{balasan}) hold, any initial
probability of the two-particle system relaxes to the time-independent
Gaussian stationary probability (\ref{yara}) \cite{risken}. The averages $\langle
x_1^2\rangle$, $\langle x_1 x_2\rangle$ and $\langle x_1^2\rangle$ are
deduced from the stationarity equation $\partial_1 J_1(\x)+\partial_2
J_2(\x)=0$:
\BEA
\label{om1}
&&\langle x_1^2\rangle = \frac{a_2T_1(a_1\gamma_1+a_2\gamma_2)+b^2 \gamma_1 (T_2-T_1)}{(a_1a_2-b^2)(a_1\gamma_1+a_2\gamma_2)},\\
\label{om2}
&&\langle x_2^2\rangle = \frac{a_1T_2(a_1\gamma_1+a_2\gamma_2)+b^2 \gamma_2 (T_1-T_2)}{(a_1a_2-b^2)(a_1\gamma_1+a_2\gamma_2)},\\
\label{om3}
&&\langle x_1 x_2\rangle = -\frac{b(a_2\gamma_2T_1+a_1\gamma_1T_2)}{(a_1a_2-b^2)(a_1\gamma_1+a_2\gamma_2)},\\
\label{om4}
&&\langle x_1^2\rangle\langle x_2^2\rangle-\langle x_1 x_2\rangle^2 \nonumber\\
&&~~~~~~=\frac{b^2\gamma_1\gamma_2(T_1-T_2)^2+(a_1\gamma_1+a_2\gamma_2)^2T_1T_2}
{(a_1a_2-b^2)(a_1\gamma_1+a_2\gamma_2)^2}, ~~~~~~
\EEA
where 
\BEA
\gamma_1\equiv {1}/{\Gamma_1}, \quad \gamma_2\equiv {1}/{\Gamma_2},
\nn
\EEA
and where the conditions $\langle x_1\rangle=\langle x_2\rangle=0$ hold automatically.
Let us introduce the following dimensionless parameters:
\BEA
\label{praha}
\varphi=\frac{\gamma_1a_1}{\gamma_2a_2}=\frac{\Gamma_2a_1}{\Gamma_1a_2}, \quad \xi=\frac{T_1}{T_2},
\quad \kappa=\frac{b^2}{a_1a_2},
\EEA
where $\varphi$ is the ratio of time-scales, while $\kappa$ characterizes the interaction strength; note
that $0<\kappa<1$ due to (\ref{balasan}).

Using (\ref{om1}--\ref{praha}) together with
(\ref{kob1}) and (\ref{kob2}), we get for the information flow,
the total entropy production, and the efficiency
\BEA
\label{damka1}
&& \i_{2\to 1}= \frac{a_2}{\Gamma_2}\,\,\frac{\kappa\,\varphi\,(1-\xi)\,(\xi+\varphi)}{\kappa\,\varphi(1-\xi)^2
    +\xi(1+\varphi)^2}, \\
\label{damka2}
&& \ell=\frac{a_2}{\Gamma_2}\,\,\frac{ (1-\xi)^2}{\xi}\,
   \frac{\kappa\,\varphi}{1+\varphi},\\
\label{damka3}
&& \eta_{2\to 1}=\frac{\xi}{1-\xi}\,\, 
   \frac{(\xi+\varphi)(1+\varphi)}{ \kappa \varphi(1-\xi)^2 +\xi (1+\varphi)^2  }.
\EEA

Let us for simplicity assume that $\xi<1$, i.e., $T_2>T_1$.  Now
$\i_{2\to 1}>0$. It is seen that both $\i_{2\to 1}$ and $\ell$ are
monotonically increasing function of $\varphi$. Thus both the
information transfer and the overall entropy production maximize when
the system attached to the hotter bath at the temperature $T_2$ is
slower [the same holds for the overall heat dissipation]. The
efficiency $\eta_{2\to 1} =\i_{2\to 1}/\ell$ is also equal to its
maximal value (\ref{fox}) in the same limit $\varphi\to \infty$; 
see (\ref{damka3}). 

The behaviour with respect to the dimensionless coupling $\kappa$ is
different.  Now $\i_{2\to 1}$ and $\ell$ are again monotonically
increasing functions of $\kappa$.  They both maximize for $\kappa\to 1$,
which|as seen from (\ref{om1}, \ref{om2}, \ref{om3})|means at the
instability threshold, where the fluctuations are very large.  However,
the efficiency $\eta_{2\to 1}$ is a monotonically decreasing function of
$\kappa$.  It maximizes (as a function of $\kappa$) in the almost
uncoupled limit $\kappa\to 0$. Thus, as far as the behaviour with
respect of the coupling constant is concerned, there is a
complementarity between maximizing the information flow and maximizing
its efficiency. 

\comment{ \BEA
\label{atas1}
&&\i_{2\to 1}= \frac{b^2 \gamma_1\gamma_2(T_2-T_1)(a_2\gamma_2T_1+a_1\gamma_1T_2)}
             {b^2 \gamma_1\gamma_2(T_2-T_1)^2+(a_2\gamma_2T_1+a_1\gamma_1T_2)^2T_1T_2},\nn\\
\label{atas2}
&&\ell=\ell_1+\ell_2=\frac{(T_1-T_2)^2}{T_1T_2}\,\frac{b^2\gamma_1\gamma_2}{a_1\gamma_1+a_2\gamma_2},\nn\\
&& T_1\ell_1+T_2\ell_2=\frac{b^2 \gamma_1\gamma_2(T_1-T_2)^2(a_2\gamma_2T_1+a_1\gamma_1T_2) }
{T_1T_2(a_1\gamma_1+a_2\gamma_2)^2  +b^2 \gamma_1\gamma_2 (T_1-T_2)^2}.
\nonumber\\
\label{atas3}
\EEA In terms of these parameters we get }

\subsection{Non-stationary situation.}
\label{non-stato}

Once the upper bound (\ref{wolf}) on the efficiency of information
transfer is established in the stationary situation, we
ask whether it might survive also in the non-stationary case.  Below we 
reconsider the coupled harmonic oscillators and show that, although the upper 
limit (\ref{wolf}) can be exceeded by some non-stationary states,
the efficiencies $\eta_{2\to 1}$ and $\eta_{1\to 2}$ are still limited
from above. 

We assume that a non-equilibrium probability of the two harmonic
oscillators is Gaussian, as given by (\ref{yara}, \ref{hajro}).  The
information flow and the entropy production are given by
(\ref{kob1}) and (\ref{kob2}), respectively. Any Gaussian probability can play the
role of some non-stationary state, which is chosen, e.g., as an initial
condition.  

To search for an upper bound of the efficiency in the non-stationary
situation, we assume a Gaussian probability (\ref{yara}) and maximize
the efficiency (\ref{dalila}) over the Hamiltonian, i.e., over the
parameters $a_1$, $a_2$ and $b$ in (\ref{hamo}).  The stability
conditions (\ref{balasan}), which are necessary for the existence of the
stationary state, are not necessary for studying non-stationary
situations (imagine a non-stationary down-hill moving oscillator).  Thus
conditions (\ref{balasan}) will not be imposed during the maximization
over $a_1$, $a_2$ and $b$. 

Note that $\i_{2\to 1}$ in (\ref{kob1}) does not depend on $a_1$ and
$a_2$. Since we are interested in maximizing the efficiency (\ref{dalila}), the first step in
this maximization is to minimize the local entropy production $\ell_1$
[given by (\ref{kob2})] over $a_1$. This produces
\BEA
a_1=\frac{T_1-b\sigma_{12}}{\sigma_{11}}, \quad \ell_1=\frac{T_1}{\Gamma_1}\,
\frac{(\beta_1 b-B)^2}{A_2}.
\label{bruno}
\EEA
The second entropy production $\ell_2$ is analogously minized over $a_2$. 
The resulting expression for the efficiency reads:
\BEA
\eta_{2\to 1}=\frac{\sigma_{12}}{\Gamma_1\sigma_{11}}\,
\frac{T_1B-b}{\frac{\beta_1}{\Gamma_1}\,
\frac{(T_1B-b)^2}{A_2}+\frac{\beta_2}{\Gamma_2}\,
\frac{(T_2B-b)^2}{A_1}}.
\label{bund}
\EEA
Next, we maximize this expression over $b$. Let us for simplicity assume 
$\sigma_{12}>0$, since the conclusions do not change for $\sigma_{12}<0$. 
The maximal value of $\eta_{2\to 1}$ reads
\BEA
\eta_{2\to 1}=
\frac{T_1}{2|T_2-T_1|}\,[\, {\rm sign}(T_2-T_1)+\sqrt{1+\chi} \,],
\label{bard}
\EEA
where
\BEA
\chi\equiv\frac{T_2\Gamma_2A_1}{T_1\Gamma_1A_2}=\frac{T_2\Gamma_2\sigma_{22}}{T_1\Gamma_1\sigma_{11}},
\EEA
and where the RHS of (\ref{bard}) is reached for
\BEA
\label{kk1}
b=-\frac{\sigma_{12}}{\sigma_{11}\sigma_{22}-\sigma_{12}^2}\,\left(T_1+\frac{|T_2-T_1|}{\sqrt{1+\chi}}
\right).
\EEA
The information flow $\i_{2\to 1}$ at the optimal value of $b$ reads
\BEA
\label{comba}
\i_{2\to 1} 
= \frac{\sigma_{12}^2}{\sigma_{11}}\, \frac{|T_1-T_2|}{\sqrt{1+\chi}}. 
\EEA

Note that the maximal value of $\eta_{2\to 1}$ does not require a strong
inter-particle coupling.  This coupling, as quantified by (\ref{kk1})
can be small due to $\sigma_{12}\to 0$. 

Now we assume that the temperatures $T_1$ and $T_2$ are fixed. It is
seen from (\ref{bard}) that taking $\chi$ sufficiently large|either due
to a large $\frac{\Gamma_2}{\Gamma_1}$, which means that the second
particle is slow, or due to a large $\frac{\sigma_{22}}{\sigma_{11}}$,
which means that its dispersion is larger|we can achieve efficiencies as
large as desired. In all these cases, for a sufficiently large $\chi$,
$\eta_{2\to 1}$ scales as $\sqrt{\chi}$.  Thus one can overcome the
stationary bound (\ref{wolf}) via some special non-stationary states and
the corresponding Hamiltonians. We, however, see from (\ref{comba}) that
increasing the efficiency due to $\chi\to\infty$ leads to decreasing the information
flow. This is the same trend as in the stationary case; see (\ref{wolf}). 
 
The large values of $\eta_{2\to 1}$ do not imply anything special for
the efficiency of the inverse information transfer efficiency
$\eta_{1\to 2}$. Indeed, the partially optimized $\eta_{1\to 2}$ reads
analogously to (\ref{bund})
\BEA
\eta_{1\to 2}=\frac{\sigma_{12}}{\Gamma_2\sigma_{22}}\,
\frac{T_2B-b}{\frac{\beta_1}{\Gamma_1}\,
\frac{(T_1B-b)^2}{A_2}+\frac{\beta_2}{\Gamma_2}\,
\frac{(T_2B-b)^2}{A_1}}.
\label{bund2}
\EEA
At the values (\ref{bruno}, \ref{kk1}), where $\eta_{2\to 1}$ extremizes, 
$\eta_{1\to 2}$ assumes a simple form
\BEA
\eta_{1\to 2} = \frac{T_2}{2(T_1-T_2)},
\nn
\EEA
i.e., $\eta_{1\to 2}$ depends only on the temperatures and can have
either sign, depending on the sign of $T_1-T_2$. Note that $\eta_{1\to 2}$
follows the same logics as in the stationary state: it is positive for $T_1>T_2$.

\section{Complementarity between heat- and information-flow}
\label{complo}

Contrary to the notion of information flow, the heat flow from one
brownian particle to another calls for an additional discussion. The
main reason for this is that the brownian particles could, in general,
be non-weakly coupled to each other, and then the local energy of a
single particle is not well defined; for various opinions on this point
see \cite{cook,segal}. In contrast, the entropy of a single particle is
always well-defined; recall in this context feature {\bf 3} in section
\ref{basso}.  Nevertheless, the notion of a separate energy can be
applied once there are physical reasons for selecting a particular form
of the interaction Hamiltonian $H_{12}(x_1,x_2)$ in (\ref{aka}), and the
average value of $H_{12}(x_1,x_2)$ is conserved in time, at least
approximately.  (A particular case of this is when $H_{12}(x_1,x_2)$ is
small.) Now $H_1(x_1)$ can be defined as the local energy of the first
particle, while the average energy change is
\BEA
\frac{\d}{\d t}\left[\int \d x_1 
P_1(x_1)H_1(x_1)\right]\equiv \frac{\d E_1}{\d t}.  
\nn
\EEA
The energy flow $\e_{2\to 1}$ from the second particle to the first one is
defined in full analogy with (\ref{muter})
\BEA
\e_{2\to 1} =\frac{\d E_1}{\d t}-\frac{\dbarrm _1 Q}{\d t},
\label{calvin}
\EEA
where $\frac{\dbarrm _1 Q}{\d t}$ is defined in (\ref{u1}). In words
(\ref{calvin}) means: the energy change $\frac{\d E_1}{\d t}$ of the
first particle is equal to the energy $\e_{2\to 1}$ received from the
second particle plus the energy $\dbarrm _1 Q$ put into the system via
coupling of the first particle to its thermal bath.

Working out (\ref{calvin}) we obtain
\BEA
\e_{2\to 1} =-\int \d \x J_1(\x;t)\partial_1 H_{12}(\x).
\label{jack}
\EEA
The interpretation of (\ref{jack}) is straightforward via concepts
introduced in Appendix \ref{coarse-grained}, where we argue that
$v_1(\x;t)=\frac{J_1(\x;t)}{P(\x;t)}$ can be regarded as a
coarse-grained velocity of the particle $1$.  The (\ref{jack}) becomes
the average work done on the particle $1$ by the force $\partial_1
H_{12}(\x)$ generated the particle $2$. 

The symmetrized heat flow is equal to the minus change of the interaction energy
\BEA
\label{akkerman}
\e_{2\to 1}+\e_{1\to 2}=-\frac{\d}{\d t}\int\d \x\, H_{12}(x) \, P(\x;t).
\EEA
Thus the there is a mismatch between the heat flowing from 1 to 2, as
compared to the energy flowing from 2 to 1. This mismatch is driven by
the change of interaction energy, and it is small provided that the
average interaction energy is [approximately] conserved in time.

It is important to stress that the ambiguity in the definition of heat
flow, which was related to the choice of the local energy $E_1$, is
absent in the stationary state, since the the average interaction energy
is conserved by construction. Now the choice of the single-particle
energy is irrelevant, since any such definition will lead to
time-independent quantity, which would then disappear from
(\ref{calvin}), $\frac{\d E_1}{\d t}=0$, and from the RHS of
(\ref{akkerman}). 

As clear from its physical meaning, the efficiency $\zeta_{2\to 1}$
of heat flow is to defined as the ratio of the heat flow over 
the total heat dissipation
\BEA
\zeta_{2\to 1}=\frac{\e_{2\to 1}}{T_1\ell_1+T_2\ell_2}.
\EEA

Let us work out $\zeta_{2\to 1}$ for the two-temperature stationary case.
Analogously to (\ref{bala1}, \ref{bala2}) we obtain
\BEA
\label{bala3}
\zeta_{2\to 1}&=&\frac{T_1}{T_2-T_1}+\frac{T_1\ell_1}{T_1\ell_1+T_2\ell_2}\\
             &=&\frac{T_2}{T_2-T_1}-\frac{T_2\ell_2}{T_1\ell_1+T_2\ell_2}.
\label{bala4}
\EEA
Eq.~(\ref{bala3}) shows that, as expected, the heat flows from higher to lower temperatures. 
In addition, (\ref{bala3}, \ref{bala4}) imply the same bounds as in (\ref{wolf}):
\BEA
\frac{T_1}{T_2-T_1}\leq \zeta_{2\to 1} \leq \frac{T_2}{T_2-T_1}.
\label{wolf1}
\EEA
Moreover, we note that the stationary efficiencies of information flow and heat flow
are related as
\BEA
\label{gossip}
\zeta_{2\to 1}\eta_{2\to 1}=\frac{T_1T_2}{(T_2-T_1)^2}.
\EEA
Note that the efficiencies $\e_{2\to 1}$ and $\eta_{2\to 1}$ in
(\ref{gossip}) depend in general on the inter-particle and
intra-particle potentials, temperatures, damping constants, {\it etc}.
However, their product in the stationary state is universal, i.e., it
depends only on the temperatures. 

Eq.~(\ref{gossip}) implies the following complementarity: for fixed
temperatures the set-up most efficient for the information transfer is
the least efficient for the heat transfer and {\it vice versa}.
Recalling our discussion in section \ref{richting} we see that the
upper bound $\frac{T_2}{T_2-T_1}$ for the efficiency $\zeta_{2\to 1}$ is
achieved for the adiabatic stationary state, where|on the contrary to
the information transfer| the hotter system is faster, i.e., $\ell_2\to
0$, but $\ell_1$ is finite. 

Note from (\ref{190}, \ref{191}) the following relation between between
the heat flow (\ref{jack}) and the information flow (\ref{soso3}):
\BEA
T_1 \i_{2\to 1} =\e_{2\to 1},
\nn
\EEA
which is valid in the adiabatic stationary state. Recall that $T_1$ here is the temperature of the bath interactng
with the fast particle. 

Let us illustrate the obtained results with the exactly solvable situation of two coupled
harmonic oscillators; see section \ref{harmo}. Recalling (\ref{hamo}--\ref{hajro}) and 
(\ref{jack}) we obtain
\BEA
\e_{2\to 1} 
= \frac{a_2 T_2}{\Gamma_2 }\, \kappa \varphi\, \frac{1-\xi}{1+\varphi},
\EEA
where we employed the dimensionless variables (\ref{praha}). 
Likewise, we get for the heat dissipation 
and the efficiency of the heat transfer
[see (\ref{damka1}--\ref{damka3}) for similar formulas]
\BEA
&& T_1\ell_1+T_2\ell_2 = \frac{a_2 T_2}{\Gamma_2 }\, \frac{\kappa\varphi(1-\xi)^2(\xi+\varphi)}{\xi(1+\varphi)^2+\kappa\varphi (1-\xi)^2},~~~~~~~~\\
&& \zeta_{2\to 1} = \frac{\xi(1+\varphi)^2+\kappa\varphi (1-\xi)^2}{(1+\varphi)(\xi+\varphi)(1-\xi)}.
\EEA
It is seen that the efficiency $\zeta_{2\to 1}$ maximizes at the
instability threshold $\kappa\to 1$, in contrast to the efficiency
$\eta_{2\to 1}$ of information transfer that maximizes at the weakest
interaction; see section \ref{harmo}. As we already discussed above, as
a function of the time-scale $\varphi$, $\zeta_{2\to 1}$ maximizes for
$\varphi\to 0$, again in contrast to the behaviour of $\eta_{2\to 1}$. 
\comment{
In analogy to the analysis presented in section \ref{non-stato} one can
show that the upper bound $\frac{T_2}{T_2-T_1}$ for the efficiency
$\zeta_{2\to 1}$ does not hold for certain non-stationary states.
However, and again analogously to the information flow efficiency $\eta_{2\to 1}$, 
the efficiency of heat flow is always bounded with the parameters of the system
(e.g., temperatures $T_1$, $T_2$ and the constants $\Gamma_1$ and $\Gamma_2$).}

\section{Transfer Entropy}
\label{i_flow_t_entropy}

\subsection{Definition}

We consider the task of predicting the future of $X_1$ from its own past
(with or without the help of the present $X_2$). Quantifying this task
leads to a concept, which has been termed directed transinformation
\cite{marko} or transfer entropy \cite{schreiber}. In the following we
will use the latter name, as this seems to be more broadly accepted.
Closely related ideas were expressed in \cite{tsuda}. The notion of
transfer entropy became recently popular among researchers working in
various inter-disciplinary fields; see \cite{prokop} for a short review. 

Our discussion of transfer entropy aims at two purposes. First, the
information flow and transfer entropy are two different notions, and their
specific differences should be clearly understood, so as to avoid any
confusion \cite{nicols}. Nevertheless, in one particular, but important case,
we found interesting relations between these two notions. 

To introduce the idea of transfer entropy let us for the moment assume
that the random quantities $X_1(t)$ and $X_2(t)$ (whose realizations are
respectively the coordinates $x_1$ and $x_2$ of the brownian particles)
assume discrete values and change at discrete instances of time: $t,\,
t+\tau,\, t+2\tau,\ldots$. 
Recalling our discussion in section \ref{basic}, we see that the conditional entropy 
$S[X_1(t+\tau)|X_1(t)]$ is the entropy reduction (residual uncertainty) of 
$X_1(t+\tau)$ due to knowing $X_1(t)$. Likewise, 
$S[X_1(t+\tau)|X_1(t),X_2(t)]$
characterizes the uncertainty of $X_1(t+\tau)$ given both $X_1(t)$ and $X_2(t)$. The difference
$\m_{2\to 1}$ is the transfer entropy:
\begin{gather}
\m_{2\to 1} \nonumber\\
\equiv\frac{1}{\tau}\left(\, S[X_1(t+\tau)|X_1(t)]-S[X_1(t+\tau)|X_1(t),X_2(t)]\,\right) \nonumber\\
\equiv \frac{1}{\tau}(I[X_1(t+\tau):X_1(t), X_2(t)]-I[X_1(t+\tau):X_1(t)]) \nonumber\\
= \frac{1}{\tau}\,\ssum_{x_1,y_1,y_2} p(y_1, y_2;t)p(x_1; t+\tau|y_1, y_2;t)\,\,\times\nonumber\\
~~~~~~~~~~~~~~~\ln \frac{p(x_1; t+\tau|y_1, y_2;t)}{p_1(x_1; t+\tau|y_1;t)}.
\label{aberdeen}
\end{gather}
$\m_{2\to 1}$ measures the difference between predicting the future of
$X_2$ from the present for both $X_1$ and $X_2$ (quantified by $I[X_1(t+\tau)|X_1(t),
X_2(t)]$) and predicting the future of $X_2$ from its own present only
(quantified by $I[X_1(t+\tau)|X_1(t)]$). 
Note that $\m_{2\to 1}$ is always positive, since additional conditiong decreases
the entropy.
$\m_{2\to 1}$ is also equal to the mutual information
$I[X_1(t+\tau):X_2(t)|X_1(t)]$ shared between the present state of $X_2$
and the future state of $X_1$ conditioned upon the present state of
$X_1$. 

To explain the transfer entropy {\it versus} information flow, we consider again the discretized version in
Fig.~\ref{Gr} (where for simplicity we take $\tau=1$). By  standard properties of  mutual information \cite{cover}, we have
\begin{eqnarray}
\label{khan}
&&I[X_1(t+1):X_2(t)]\leq I[X_1(t+1),X_1(t):X_2(t)]~~~~~~~~\\
    &&= I[X_1(t):X_2(t)]+I[X_1(t+1):X_2(t)|X_1(t)]\, ,
\label{batu}
\end{eqnarray}
where the inequality in (\ref{khan}) is related to the strong sub-additivity feature, and where the equality
in (\ref{batu}) is the chain rule for the mutual information \cite{cover}. 
Hence,
\BEA
\i_{2\to 1}=I[X_1(t+1):X_2(t)]-I[X_1(t):X_2(t)]~~~~~~~~~~\nn\\ 
\leq I[X_1(t+1):X_2(t)|X_1(t)]=\m_{2\to 1}\,.\nn
\EEA
The LHS (left hand side) is the discrete version of information flow, the RHS
the transfer entropy. 
If the arrow from  $X_2(t)$ to $X_1(t+1)$ was absent, the 
modified graph would impose \cite{Pearl}
that $X_1(t+1)$ and $X_2(t)$ are conditionally independent, given $X_1(t)$, i.e.,
\[
\m_{2\to 1}=I[X_1(t+1):X_2(t)|X_1(t)]=0\,.
\]
This shows that the transfer entropy vanishes in this case, as it should be because there
is no arrow transmitting information. Thus, $\m_{2\to 1}$ characterizes the strength of that arrow.

For continiuous time [but still discrete variables $X_1$ and $X_2$] we take $\tau\to 0$ in (\ref{aberdeen})
producing 
\BEA
\m_{2\to 1} =\ssum_{y_2, x_1\not =y_1} p(y_1, y_2;t)g_1(x_1|y_1; y_2)\,\,\times\nonumber\\
\ln \frac{g_1(x_1|y_1; y_2)}{\ssum_{z_2} g_1(x_1|y_1; z_2)  P_{2|1}(z_2|y_1;t)},
\label{aharon}
\EEA
where $g_1(x_1|y_1; y_2)$ is defined analogously to $G_1$ in (\ref{veda_1}), but with discrete random variables.

Eq.~(\ref{aharon}) cannot be translated to the continuous variable
situation simply by interchanging the probabilities $p$ with the
probability densities $P$, since attempting such a translation leads to
singularities. The proper extension of (\ref{aberdeen}) to continuous
variables and continuous time reads
\begin{gather}
\m_{2\to 1}={\rm lim}_{\tau\to 0}\,\frac{1}{\tau}
\int \d \y P(\y;t)\int \CP[\d x_1\,^{t+\tau}_t |\y;t]\,\times \nonumber\\
\ln \frac{\CP[\d x_1\,^{t+\tau}_t |\y;t]}{\CP[\d x_1\,^{t+\tau}_t |y_1;t]},
\label{celtic}
\end{gather}
where $\CP[\d x_1\,^{t+\tau}_t |\y;t]$ ($\CP[\d x_1\,^{t+\tau}_t |y_1;t]$) 
is the measure of all paths $x_1(t)$ starting from $\y=(y_1,y_2)$ (from $y_1$) at time $t$ and ending
somewhere at time $t+\tau$, i.e., not the final point of the path is fixed, but rather the initial time and final times
\footnote{Integrating $\CP[\d x_1\,^{t+\tau}_t |\y;t]$ over all paths starting from $\y$ at time $t$ and ending at $x_1$ in
time $t+\tau$ we get the conditional probability density $P(x_1,t+\tau|\y,t)$ \cite{larry}.}.
This extension naturally 
follows general ideas of information theory in continuous spaces \cite{strat_IT}.
For our situation the measures $\CP[\d x_1\,^{t+\tau}_t |\y;t]$ and $\CP[\d x_1\,^{t+\tau}_t |y_1;t]$
refer to the stochastic process described by 
(\ref{1}, \ref{160}); see \cite{larry} for an introduction to such measures. 

Eq.~(\ref{celtic}) is worked out in Appendix \ref{schwab} producing
\BEA
\label{glan}
\m_{2\to 1}=\frac{1}{2T_1\Gamma_1}\int \d \x\, P(\x;t)\, \phi_{2\to 1}^2(\x;t),~~~~~~~~~~~~~~~\\
\phi_{2\to 1}\equiv
\pd_1 H_{12}(\x)-\int\d y_2 \,\pd_1 H_{12}(x_1,y_2)P_{2|1}(y_2|x_1;t),
\nn
\EEA
where $\phi_{2\to 1}$ is the force acting from $2$ to $1$ minus its conditional average; compare with (\ref{191}).

\subsection{Information flow versus entropy transfer: General differences.}

Let us compare features of the entropy transfer $\m_{2\to 1}$ to those
of information flow $\i_{2\to 1}$.  We remind that difference between
the information flow $\i_{2\to 1}$ and transfer entropy $\m_{2\to 1}$
stem from the fact that $\m_{2\to 1}$ refers to the prediction of the
future of $X_2$ from its own past (with or without the help of the
present of $X_1$), while $\i_{2\to 1}$ refers to the prediction gain (or
loss) of the future of $X_2$ from the present of $X_1$. Thus for
$\m_{2\to 1}$ the active agent is $1$ predicting its own future, while for
$\i_{2\to 1}$ the active agent is $2$ prediciting the future of $1$.

{\it i)} Both $\m_{2\to 1}$ and $\i_{2\to 1}$ are invariant with respect
to redefining the interaction Hamiltonian; see (\ref{hoviv}) and compare
with information flow $\i_{2\to 1}$. 

{\it ii)} In contrast to the information flow
$\i_{2\to 1}$, the entropy transfer $\m_{2\to 1}$ is always
non-negative. 

{\it iii)} In contrast to $\i_{2\to 1}$, $\m_{2\to 1}$ does not nullify
for factorized probabilities $P(\x)=P_1(x_1)P_2(x_2)$, provided that
there is a non-trivial interaction $H_{12}$. This because $\m_{2\to 1}$
is defined with respect to the transition probabilities; see
(\ref{aberdeen}). 

{\it iv)} $\m_{2\to 1}$ nullifies whenever there is no force acting from
one particle to another. Recall that the force-driven part $\i^{\rm
F}_{2\to 1}$ of the information flow also nullifies together with the
force, albeit $\i^{\rm F}_{2\to 1}$ nullifies also for factorized
probabilities; see (\ref{hoviv}).

{\it v)} In contrast to $\i_{2\to 1}$, $\m_{2\to 1}$ does not nullify at
equilibrium.  Thus, $2$ can help $1$ in predicting its future at
absolutely no thermodynamic cost.  However, as we have shown, there is a
definite thermodynamic cost for $2$ wanting to predict the future of $1$
better than it predicts the present of $1$.

{\it vi)} $\m_{2\to 1}$ is not a flow, since it does not add up
additively to time-derivative of any global quantity.  However,
obviously $\m_{2\to 1}$ does refer to some type of information
processing. In fact, $\m_{2\to 1}$ underlies the notion of Granger-causality, which was first proposed in the
context of econometrics
\cite{granger_general_definition,granger_causality} (see
\cite{palus_review} for a review): the ratio $\frac{\m_{2\to
1}}{\m_{1\to 2}}$ quantifies the strength of causal influences from $2$
to $1$ {\it relative} to those from $1$ to $2$ \footnote{The usage of
ratio $\frac{\m_{2\to 1}}{\m_{1\to 2}}$ is obligatory, since $\m_{2\to
1}$ does not characterize the absolute strength of the influence of $2$
on $1$. Note, e.g., that $\m_{2\to 1}$ nullifies not only for
independent process $X_1(t)$ and $X_2(t)$, but also for identical
(strongly coupled) processes $X_1(t)=X_2(t)$; see (\ref{aberdeen}). This
point is made in \cite{nicols}.  }.  
The notion of Granger-causality is
useful as witnessed by its successful empirical applications
\cite{nicols,prokop,palus_review,granger_general_definition,granger_causality}.
For $\frac{\m_{2\to 1}}{\m_{1\to 2}}\gg 1$ we shall tell that $2$ is
Granger-driving $1$. 

\comment{As any other attempt of catching causal relations without doing
direct interventions (it may be difficult or even impossible to
intervene in practice), the notion of Granger-causality is certainly
limited
\cite{nicols,prokop,palus_review,granger_general_definition,granger_causality}.
This is why the very term Granger-causality emerged to distinguish it
from the "real" causality.}

\subsection{Information flow versus transfer entropy in the adiabatic stationary limit}

Given the differences between the information flow and the entropy
transfer it is curious to note that in the adiabatic stationary
situation (see section \ref{aadiaa}) there exist a direct relation
between them. We recall that the adiabatic situation is special, since
the eficiency of information flow reaches its maximal value there.
Recall that this situation is defined (besides the long-time limit) by
condition (\ref{astrakhan}), which means that $2$ is slow, while $1$ is
fast; at equilibrium, when $T_1=T_2$, this slow versus fast separation
becomes irrelevant. Reminding also the definition (\ref{hoviv}) of the
force-driven part $\i^{\rm F}_{2\to 1}$ of the information flow, we see that
\BEA
\frac{\m_{2\to 1}}{\m_{1\to 2}}={\cal O}\left(\frac{\Gamma_2}{\Gamma_1}\right)\gg 1,
\quad 
\frac{\i^{\rm F}_{2\to 1}}{\i^{\rm F}_{1\to 2}}={\cal
O}\left(\frac{\Gamma_2}{\Gamma_1}\right)\gg 1. 
\label{ola}
\EEA
The first relation in (\ref{ola}) indicates that the slow system is Granger-driving the fast one, while
the second relation implies that the same qualitative conclusion is got from looking at 
$\i^{\rm F}_{2\to 1}$. This point is strengthened by noting from (\ref{osh}, \ref{glan}) that in 
in the adiabatic, stationary, two-temperature situation we have
\BEA
\label{akh1}
\m_{2\to 1}=\frac{1}{2}\,\i^{\rm F}_{2\to 1}. 
\EEA
A less straighforward relation holds for the action of the fast system on the slow one
\BEA
\label{akh2}
\m_{1\to 2}=\frac{T_1}{T_2}\,\frac{1}{2}i^{\rm F}_{1\to 2}.
\EEA
Recall that for the considered adiabatic stationary state, it is the action of the fast on the slow that 
determines the magnitude of the information flow $\i_{1\to 2}$ (the sign of $\i_{1\to 2}$ is fixed
by the temperature difference):
\BEA
\label{akh3}
\i_{1\to 2}=-\i_{2\to 1}
=\frac{T_1-T_2}{T_1}\,\i^{\rm F}_{1\to 2}.
\EEA
It is seen that provided $T_2>T_1$ (i.e., the slow system is attached to
the hotter bath, a situation realized for the optimal information
transfer) the Granger-driving qualitatively coincides with the causality
intuition implied by the sign of information flow: both $\i_{2\to 1}>0$
and $\frac{\m_{2\to 1}}{\m_{1\to 2}}\gg 1$ hold, which means that $2$
predicts better the future of $1$ ($\i_{2\to 1}>0$) and that $2$ is more
relevant for helping $1$ to predict its own future ($\frac{\m_{2\to
1}}{\m_{1\to 2}}\gg 1$). 

It is tempting to suggest that only when the causality intuition deduced
from $\i_{2\to 1}$ agrees with that deduced from $\m_{2\to 1}$, we are
closer to gain a real understanding of causality (still without doing
actual interventions). Interestingly, the present slow-fast
two-temperature adiabatic system was considered recently from the
viewpoint of other non-interventional causality detection methods
reaching a similar conclusion: unambiguous causality can be detected in
this system, if the slow variable is attached to the hot thermal bath
\cite{Domini}. 

For the harmonic-oscillator example treated in 
section \ref{harmo} we obtain
\BEA
\m_{2\to 1}=\frac{b^2}{2A_2T_1\Gamma_1}, \quad 
\i^{\rm F}_{2\to 1}=\frac{bB}{A_2\Gamma_1}.
\EEA
Employing formulas (\ref{yara}, \ref{hajro}) and (\ref{om1}--\ref{om4}) 
for the stationary state we note the following
relation
\BEA
\label{bonn}
\frac{\i^{\rm F}_{2\to 1}}{\i^{\rm F}_{1\to 2}}&=&\frac{T_1}{T_2}\,\, \frac{\m_{2\to 1}}{\m_{1\to 2}}\\
&=& \frac{1}{\xi} \, \frac{\kappa (\xi-1)+\varphi+1}{\kappa (\frac{1}{\xi}-1)+\frac{1}{\varphi}+1} ,
\label{marburg}
\EEA
where the dimensionless parameters $\kappa$, $\xi$ and $\varphi$ are
defined in (\ref{praha}).  In the adiabatic situation (\ref{bonn}) is
naturally consistent with (\ref{akh2}, \ref{akh3}), but for the
considered harmonic ocillators it is valid more generally, i.e., for an
arbitraty stationary state. Eq.~(\ref{marburg}) explicitly demonstrates the
conflict in Granger-driving between making the oscillator $2$ slow (i.e., $\varphi\to\infty$) and making
it cold (i.e., $\xi\to\infty$): for $\varphi\to\infty$, $\frac{\m_{2\to 1}}{\m_{1\to 2}}$
tends to infinity, while for $\xi\to\infty$, $\frac{\m_{2\to 1}}{\m_{1\to 2}}$ tends to zero.

Note finally that
\BEA
\frac{\i^{\rm F}_{2\to 1}}{\m_{2\to 1}}=\frac{\xi (1+\varphi)(\xi+\varphi)}{\kappa\varphi+\xi(1+\varphi)^2},
\nn
\EEA
which means that apart from the adiabatic limit ($\varphi\to 0$ or $\varphi\to \infty$)
there is no straightforward relation between $\i^{\rm F}_{2\to 1}$ and $\m_{2\to 1}$.

\section{Summary}
\label{summo}

We have investigated the task of information transfer implemented on a
special bi-partite physical system (pair of Brownian particles, each
coupled to a bath). Our main conclusions are as follows:

{\bf 0.} The information flow $\i_{2\to 1}$ from one Brownian particle
to another is defined via the time-shifted mutual information. For the
considered class of systems this definition coincides with the entropy
flow, as defined in statistical thermodynamics. 

{\bf 1.} The information flow $\i_{2\to 1}$ is a sum of two terms:
$\i_{2\to 1}=\i^{\rm F}_{2\to 1}+\i^{\rm B}_{2\to 1}$, where the bath
driven contribution $\i^{\rm B}_{2\to 1}\leq 0$ is the minus Fisher
information, and where the force-driven contribution $\i^{\rm F}_{2\to
1}$ has to be positive and large enough for the particle $2$ to be an
information source for the particle $1$.

{\bf 2.} No information flow from one particle to another is possible
in equilibrium. This fact is recognized in literature \cite{porod},
though by itsels it does not yet point out to a definite thermodynamic
cost for information transfer. 

{\bf 3.} For a stationary non-equilibrium state created by a finite difference
between two temperatures $T_1<T_2$, the ratio of the information flow to
the total entropy production|i.e., the efficiency of information flow|
is limited from above by $\frac{T_2}{T_2-T_1}$.  This bound for the
efficiency defines the minimal thermodynamic cost of information flow
for the studied setup. Note that not the total amount of transferred information,
but rather its rate is limited. Thus the thermodynamic cost accounts also for the time
during which the information is transferred. 

{\bf 4.} The upper bound $\frac{T_2}{T_2-T_1}$ is reachable in the
adiabatic limit, where the sub-systems have widely different
characteristic times. The information flow is then small on the
time-scale of the fast motion, but sizable on the time-scale of the slow
motion. 

{\bf 5.} The information transfer between two sub-systems (Brownian
particles) naturally nullifies, if these system are not interacting, and
were not interacting in the past. It is thus relevant to study how the
efficiency and information flow depend on the inter-particle coupling
strength.  As functions of the inter-particle coupling strength, the
efficiency and information flow demonstrate the following
complementarity. The information flow is maximized at the instability
threshold of the system (which is reached at the strongest coupling
compatible with stability). On the contrary, the efficiency is maximized
for the weakest coupling. 

{\bf 6.} There are special two-temperature, but non-stationary scenarios,
where the efficiency of information flow is much larger than
$\frac{T_2}{T_2-T_1}$, but it is still limited by the basic parameters
of the system (the ratio of the time-scales and the ratio of
temperatures).

{\bf 7.} Analogous consideration can be applied to the energy (heat)
flow from one sub-system to another. The efficiency of the heat
flow|which is defined as the heat flow over the total amount of the
heat dissipated in the overall system|is limited from above by the
same factor $\frac{T_2}{T_2-T_1}$ (assuming that $T_1<T_2$). However, in
the stationary state there is a complementarity between heat flow and
information flow: the setup which is most efficient for the information
transfer is the least efficient for the heat transfer and {\it vice
versa}. 

{\bf 8.} There are definite relations
between the information flow and the transfer entropy introduced in
\cite{marko,schreiber}. The transfer entropy is not a flow of
information, though it quantifies some type of information processing in
the system, a processing that occurs without any thermodynamic cost. 

\subsection*{Acknowledgements} The
Volkswagenstiftung is acknowledged
for financial support (grant "Quantum Thermodynamics: Energy and Information flow
at nano-scale"). A. E. A. was supported by ANSEF and SCS of Armenia (grant 08-0166).

\appendix

\section{Information-theoretic meaning of entropy and mutual information}
\label{info-meaning}

\subsection{Entropy}

Let us recall the information-theoretic meaning of entropy (\ref{entropy_discrete}), i.e., in which specific
operational sense $S[X]$ quantifies the amount of information contained
in the random variable $X$.  To this end imagine that $X$ is composition
of $N$ random variables $\{X(1),\ldots,X(N)\}$, i.e., $X$ is a 
random process. We assume that this process is ergodic \cite{ash}. The simplest
example of such a process is the case when $X(1)$, $\ldots$, $X(N)$
are all independent and identical,
\BEA
p(x(1),\ldots, x(N))=\pprod_{k=1}^N p(x(k)),
\label{garun}
\EEA
where $x(k)=1,\ldots,n$ parametrize realizations of $X(k)$.  Note
that for an ergodic process the entropy in the limit $N\gg 1$ scales as
$\propto N$, e.g., $S[X]=-N{\ssum}_{k=1}^n p(k)\ln p(k)$ for the above
example (\ref{garun}). 

For $N\gg 1$, the set of $n^N$ realizations of the ergodic process $X$
can be divided into two subsets \cite{ash,strat_IT,cover}. The first subset
$\Omega(X)$ is called typical, since this is the minimal subset with the
probability converging to $1$ for $N\gg 1$ \cite{ash,strat_IT,cover}; the
convergence is normally exponential over $N$. The number of elements in $\Omega(X)$ grows asymptotically as
$e^{S[X]}$ for $N\gg 1$. These elements have (nearly) equal probabilities $e^{-S[X]}$.
Thus the number of elements in $\Omega(X)$ is generally much smaller
than the overall number of realizations $e^{N\ln n}$.

The overall probability of those realizations which do not fall into
$\Omega(X)$, scales as $e^{-{\rm const\,}N}$, and is neglegible in the
thermodynamical limit $N\gg 1$. All these features are direct
consequences of the law of large numbers, which holds for ergodic processes
at least in its weak form \cite{ash,strat_IT,cover}. Since in the limit $N\gg
1$ the realizations of the original random variable $X$ can be in a
sense substituted by the typical set $\Omega(X)$, the number of elements
in $\Omega(X)$ characterizes the information content of $X$
\cite{ash,strat_IT,cover}.

\subsection{Mutual information}

While the entropy $S[X]$ reflects the information content of the
(noiseless) probabilistic information source, the mutual information
$I[Y:X]$ characterizes the maximal information, which can be shared
through a noisy channel, where $X$ and $Y$ correspond to the input and
output of the channel \cite{ash,strat_IT,cover}, respectively.
To understand the qualitative content of
this relation consider an ergodic process $XY=\{X(1)Y(1),\ldots, X(N)
Y(N)\}$, where
$x(l)=1,\ldots,n$ and $y(l)=1,\ldots,n$ are the realizations
of $X(l)$ and $Y(l)$, respectively.

One now looks at $X$ ($Y$) as the input (output) of a noisy
channel \cite{ash,strat_IT,cover}. 
In the limit $N\to \infty$
we can study the typical sets $\Omega(.)$ instead of the full set of realizations for the random
variables. It appears for $N\to \infty$ that the typical sets $\Omega(X)$ and 
$\Omega(Y)$ can be represented as union of $M\equiv e^{I[X:Y]}$ non-overlapping
subsets \cite{strat_IT}:
\BEA
   \Omega(X)= {\bigcup}_{\alpha=1}^M \omega_\alpha(X), ~~
   \Omega(Y)= {\bigcup}_{\alpha=1}^M \omega_\alpha(Y), 
\EEA
such that for $N\gg 1$ 
\BEA
\label{conrad_1}
p[y \in \omega_\alpha(Y)\, |\, x \in \omega_\beta(X)] 
=\delta_{\alpha\beta} e^{-S(Y|X)},\\
p[x \in \omega_\alpha(X)\, |\, y \in \omega_\beta(Y)] 
=\delta_{\alpha\beta} e^{-S(X|Y)}.
\label{conrad_2}
\EEA
Note that the number of elements in $\omega_\alpha(X)$
($\omega_\alpha(Y)$) is asymptotically $e^{S[X|Y]}$ ($e^{S[Y|X]}$).
Eqs.~(\ref{conrad_1}, \ref{conrad_2}) mean that the realizations from
$\omega_\alpha(Y)$ correlate only with those from $\omega_\alpha(X)$,
and that all realizations within $\omega_\alpha(X)$ and within
$\omega_\alpha(Y)$ are equivalent in the sense of (\ref{conrad_1},
\ref{conrad_2}). It should be clear that once the elements of
$\omega_\alpha( X)$ are completely mixed during the mapping to
$\omega_\alpha( Y)$, the only reliable way of sending information
through this noisy channel is to relate the reliably shared words to the
sets $\omega_\alpha$ \cite{strat_IT}. Since there are $e^{I[X:Y]}$ such
sets, the number of reliably shared words is limited by $e^{I[X:Y]}$.
Note that the number of elements in the typical set $\Omega_N(X)$ is
equal to $e^{S(X)}$, which in general is much larger than $e^{I[X:Y]}$.

\section{Operational definition of information flow}
\label{opera}

Here we demonstrate that the definition (\ref{main_definition}) of the
information flow is recovered from an operational approach proposed in
\cite{majda}; see \cite{liang_kleeman_prl,liang_kleeman_pd} for related
works. 

Let us imagine that at some time $\bar{t}$ we suddenly increase the
damping constant $\Gamma_2$ to some very large value. As seen from
(\ref{1}, \ref{160}, \ref{170}), this will freeze the dynamics of the
second particle, so that for $t>\bar{t}$ the joint probability
distribution $\bar{P}(\x;t)$ satisfies the following modified
Fokker-Planck equation
\BEA
\label{alister}
\partial _t \bar{P}(\x;t) &+& \partial _1 \bar{J}_1(\x;t) =0, \\
-\Gamma _1 \bar{J}_1(\x;t)&=& \bar{P}(\x;t)\partial _1 H(\x)
+ T_1 \partial _1 \bar{P}(\x;t),
\EEA
together with the boundary condition 
\BEA
\label{maclin}
\bar{P}(\x;\bar{t})=P(\x;\bar{t}),
\EEA
where $P(\x;\bar{t})$ satisfies the Fokker-Planck equation
(\ref{160}, \ref{170}).  Other methods of freezing the dynamics of
$X_2(t)$ (e.g., switching on a strong confining potential $\widetilde{H}(x_2)$ acting on $x_2$) would work for
the present purposes equally well. 
The fact of freezing should be apparent from (\ref{alister}) 
whose solution can be represented
[using also (\ref{maclin})] as
\BEA
\label{gambetta}
\bar{P}(\x;t)=\bar{P}_{1|2}(x_1|x_2;t)\,P_2(x_2;\bar{t}), \quad t\geq\bar{t}.
\EEA
Note that once $X_2$ is frozen, it becomes a random external field from the viewpoint of the dynamics of $X_1(t)$.
The entropy of a system in a random field is standardly calculated via averaging (over the field distribution) 
the entropy calculated at a fixed field: 
\BEA
\bar{S}_{1|2}\equiv -\int \d \x P(x_2;t)\bar{P}_{1|2}(x_1| x_2;t)\ln \bar{P}_{1|2}(x_1| x_2;t).\nn
\EEA

We select $t\to \bar{t}+0$ and note that the freezing does 
influence directly the marginal entropy rate of the
first particle
\BEA
\frac{\d S_1}{\d t}
&\equiv& \frac{\d}{\d t}\left[-\int \d x_1 P_1(x_1;t)\ln P_1(x_1;t)\right], 
\nn\\
\label{koba_2}
&=&\frac{\d}{\d t}\left.\left[-\int \d x_1 \bar{P}_1(x_1;t)\ln \bar{P}_1(x_1;t)\right]\right|_{\bar{t}=t},\nn
\EEA
a fact that follows from (\ref{alister}, \ref{maclin}, \ref{gambetta}).

Now we subtract from the entropy rate of the first particle
the rate of the conditional entropy $\bar{S}_{1|2}$:
\begin{gather}
\i_{2\to 1}(t)\equiv \frac{\d S_1}{\d t}
-\left.\frac{\d \bar{S}_{1|2}}{\d t}
\right|_{\bar{t}=t},
\label{boris}
\end{gather}
where the conditioning $\bar{t}=t$ is done {\it after} taking $\frac{\d}{\d t}$.
Thus, $\i_{2\to 1}(t)$ is that part of the entropy change of
$X_1$ (between $t$ and $t+\tau$), which exists due to fluctuations
of $X_2(t)$; see section \ref{mumu}.  

Employing (\ref{gambetta}) we note for the last part in (\ref{boris})
\BEA
\label{aramovich}
\left.\frac{\d \bar{S}_{1|2}}{\d t}
\right|_{\bar{t}=t} &=& \left.\frac{\d \bar{S}}{\d t}
\right|_{\bar{t}=t}\nn \\
&\equiv&\frac{\d}{\d t}\left.
\left[-\int \d \x \bar{P}(\x;t)\ln \bar{P}(\x;t)\right]
\right|_{\bar{t}=t},
\nonumber
\EEA
which means that $\i_{2\to 1}(t)$
can be defined equivalently via the total entropy rate $\frac{\d \bar{S}}{\d t}$ of the overall
system, with the second particle being frozen. 
Now using 
\BEA
&&\frac{\d S_1}{\d t}=\int \d \x \left[
\ln P_1(x_1;t)\,\right]\,\partial_1 J_1(\x;t),\nn\\
&& \left.\frac{\d \bar{S}}{\d t}\right|_{\bar{t}=t} =
\int \d \x \left[
\ln P(\x;t)\,\right]\,\partial_1 J_1(\x;t),
\nn
\EEA
we get back from (\ref{boris}) to (\ref{korkud}) confirming that both definitions are equivalent.

\section{Coarse-grained velocities for brownian particles}
\label{coarse-grained}

Consider an ensemble of all realizations of the two-particle Brownian
system which at time $t$ have a coordinate vector $\x$. For this
ensemble the average coarse-grained velocity for the particle with index
$j$ might naively be defined as:
\BEA
v_{j}(\x,t)={\rm lim}_{\ep\to 0}
\,\int\d \y \,\frac{y_j-x_j}{\ep}\,P(\y,t+\ep|\x,t).
\EEA
However, it was pointed out by Nelson \cite{nelson} that the absence
of regular trajectories enforces one to define different velocities for
different directions of time:
\BEA
\label{dish1}
v_{+,j}(\x,t)={\rm lim}_{\ep\to +0}
\,\int\d y_j\,\frac{y_j-x_j}{\ep}\,P(y_j,t+\ep|\x,t),~~\\
v_{-,j}(\x,t)={\rm lim}_{\ep\to +0}
\,\int\d y_j\,\frac{x_j-y_j}{\ep}\,P(y_j,t-\ep|\x,t).~~
\label{dish2}
\EEA
The physical meaning of these expressions is as follows:
$v_{+,j}(\x,t)$ is the average 
velocity to move anywhere starting from $(\x,t)$, whereas
$v_{-,j}(\x,t)$ is the average velocity to come from anywhere and to
arrive at $\x$ at the moment $t$.
Since these velocities are defined already in the overdamped limit,
$\ep$ is assumed to be much larger than the characteristic relaxation
time of the (real) momentum which is small in the overdamped limit.
Therefore, we call (\ref{dish1}, \ref{dish2}) coarse-grained
velocities. It is known that for the overdamped brownian motion almost
all trajectories are not smooth. This is connected to the chaotic
influences of the bath(s) which randomize the real momenta on much
smaller times, and this is also the reason for $v_{+,j}(\x,t)\not
=v_{-,j}(\x,t)$. The difference $v_{+,j}(\x,t)- v_{-,j}(\x,t)$
thus characterizes the degree of the above non-smoothness.
One now can show that \cite{nelson,akn}
\BEA
&&v_{+,j}(\x,t)=-\frac{1}{\Gamma_j}\partial_j H(\x),\nonumber\\
&&v_{-,j}(\x,t)=-\frac{1}{\Gamma_j}[\,\partial_j H(\x)+2T_j\,\partial_{j}\ln P(\x,t)\,].\nn
\label{katu}
\EEA
We now see that the probability times the {\it average coarse-grained
velocity} $\frac{1}{2}[v_{+,j}(\x,t)+v_{-,j}(\x,t)]\equiv v_j(\x,t)$
amounts to the probability current of the Fokker-Planck equation
(\ref{160}, \ref{170}):
\BEA
\label{shun}
v_j(\x,t)P(\x;t)=J_j(\x;t).\nn
\EEA
If one would take $\ep$ in (\ref{dish1}, \ref{dish2}) much smaller than
the characteristic relaxation time of the momentum |which would amount
to applying definitions (\ref{dish1}) and (\ref{dish2}) to a smoother
trajectory| then $v_{+,j}(\x,t)$ and $v_{-,j}(\x,t)$ would be equal to
each other and equal to the average momentum; see \cite{akn} for more
details.

\section{Calculation of transfer entropy}
\label{schwab}

Here we calculate the entropy transfer as defined in (\ref{celtic}). The measures entering this
equation read for a small $\tau$
\begin{gather}
\label{beduin_1}
\CP[\d x_1\,^{t+\tau}_t |y_1,y_2;t]= 
{\cal K}\,e^{\frac{-1}{2T_1\Gamma_1}\int_t^{t+\tau}\d \sigma [\Gamma_1 \dot{x}_1+\pd_1 H(x_1(\sigma), y_2)]^2 },\\
\CP[\d x_1\,^{t+\tau}_t |y_1;t]=
{\cal K}\, e^{\frac{-1}{2T_1\Gamma_1}\int_t^{t+\tau}\d \sigma [\Gamma_1 \dot{x}_1+h_1(x_1(\sigma))]^2 },
\label{beduin_2}
\end{gather}
where ${\cal K}$ is the normalization constant, and where
\BEA
h_1(x_1)\equiv \int \d y_2 \pd_1 H(x_1, y_2)P_{2|1}(y_2|x_1).
\EEA
Both time integrals 
$\int_t^{t+\tau}$ in (\ref{beduin_1}, \ref{beduin_2}) are to be 
interpreted in the Ito sense \cite{gardiner}. Due to this the normalization constants
for both path-integrals are identical. 

To understand the origin of (\ref{beduin_1}, \ref{beduin_2})
recall from (\ref{chapo}--\ref{veda_22})
that in the small-$\tau$ limit:
\BEA
P(x_1;t+\tau|y_1,y_2;t)=\delta(x_1-y_1)~~~~~~~~~~~~~~~~\nonumber\\
+\frac{\tau}{\Gamma_1}\,
\partial_{1}\left[
\delta(y_1-x_1)\partial_{1}H(x_1,y_2)+T_1\partial_{1} \delta(y_1-x_1)
\right],\\
P(x_1;t+\tau|y_1;t)=\delta(x_1-y_1)~~~~~~~~~~~~~~~~\nonumber\\
+\frac{\tau}{\Gamma_1}\,
\partial_{1}\left[
\delta(y_1-x_1)h_1(x_1)+T_1\partial_{1} \delta(y_1-x_1)
\right].
\EEA

For a small $\tau$ we get
\begin{gather}
\frac{1}{\tau}\frac{\CP[\d x_1\,^{t+\tau}_t |\y;t]}{\CP[\d x_1\,^{t+\tau}_t |y_1;t]}
= \frac{h^2_1(y_1)-[\pd_1 H(y_1,y_2)]^2}{2T_1\Gamma_1}\nonumber\\
+ \frac{h_1(y_1)-\pd_1 H(y_1,y_2)}{T_1}\,\,\frac{x_1(t+\tau)-x_1(t)}{\tau}.
\label{hska}
\end{gather}
The fact that the time-integrals were taken in the Ito sense is visible in the last term of (\ref{hska}).
Putting (\ref{hska}) into (\ref{celtic}) and noting [for a small $\tau$]
\BEA
\int \CP[\d x_1\,^{t+\tau}_t |\y;t] \, \frac{x_1(t+\tau)-x_1(t)}{\tau}=-\frac{1}{\Gamma_1}\pd_1 H(\y),\nn
\EEA
we end up at (\ref{glan}).

\end{document}